\documentclass[aps,prb,showpacs,twocolumn]{revtex4}
\usepackage{amsmath}
\usepackage{graphicx}

\begin{document}

\title{
Grand canonical Gutzwiller approximation
for magnetic inhomogeneous systems
}
\author{Noboru Fukushima}
\email{noboru@phys.sinica.edu.tw}
\affiliation{Institute of Physics, Academia Sinica, NanKang, Taipei 11529, Taiwan}

\begin{abstract}
 The Gutzwiller approximation (GA) for Gutzwiller-projected grand
 canonical wave functions with fugacity factors is investigated in
 detail.  Our systems in general contain inhomogeneity and local
 magnetic moments.  In deriving renormalization formulae, we also derive
 or estimate terms of higher powers of intersite contractions neglected
 in the conventional GA.  We examine several different constraints, {\it
 i.e.}, local/global spin-dependent/independent particle-number
 conservation.  Out of the four, the local spin-dependent constraint
 seems the most promising at present.  An improved GA derived from it
 agrees with the variational Monte Carlo method better than the
 conventional GA does.  The corrections to the conventional GA can be
 interpreted as two-site correlation including the phase difference of
 configurations.  Furthermore, projected quasi-particle excited states
 are orthogonal to each other within the GA.  Using these states,
 spectral weights are calculated.  We show that asymmetry between
 electron addition and removal spectra can appear by taking into account
 the higher powers of the intersite contractions in the case of the
 $d$-wave superconductors and the Fermi sea; the addition is smaller
 than the removal.  However, the asymmetry is quite weak especially near
 the Fermi level.  In contrast, projected $s$-wave superconductors can
 have the opposite asymmetry (addition larger than removal)
 especially near the Fermi level.
 In addition, formulae from the other three constraints are also
 derived, which may be useful depending on purposes.
\end{abstract}


\maketitle

\section{Introduction}

This paper concerns calculation of expectation values using projected
wave functions in inhomogeneous systems.
In order to study electronic systems with repulsive on-site
interactions,
Gutzwiller proposed projected wave functions\cite{Gutzwiller63} of the form
$P_{\rm G}|\Psi_0\rangle$ with the Gutzwiller projection operator,
\begin{equation}
 P_{\rm G} \equiv \prod_i (1-\hat{n}_{i\uparrow} \hat{n}_{i\downarrow})
~,
\end{equation}
to prohibit electron double occupancy on each site.
Here, $\hat{n}_{i\sigma} = c_{i\sigma}^\dagger c_{i\sigma}$ with
$c_{i\sigma}^\dagger$ ($c_{i\sigma}$) being the creation (annihilation)
operator of site $i$ and spin $\sigma$.

Expectation values of operators by this projected wave function
can be evaluated  by the variational Monte Carlo method (VMC)
numerically exactly within statistical errors.
However, the VMC requires lots of computational effort for some issues.
In addition, it needs one run for each parameter set, whereas
an analytical method can generate more general formulae that
often provide us some hint to understand the system.
Thus, instead of the VMC,
an analytical approximation called the Gutzwiller approximation (GA)
is used on occasions, {\it i.e.},
\begin{equation}
{\langle \Psi^N| \hat{O} |\Psi^N\rangle
\over
\langle  \Psi^N|\Psi^N\rangle}\
\approx \ g^{O} \,
{\langle \Psi^N_0| \hat{O} |\Psi^N_0\rangle
\over
\langle \Psi^N_0|\Psi^N_0\rangle}
~,
\label{renorm}
\end{equation}
with $|\Psi^N \rangle\equiv P_{\rm G} |\Psi^N_0 \rangle$, where
$|\Psi^N_0 \rangle$ have a fixed particle number $N$.
The factor $g^{O}$ is the Gutzwiller renormalization factor for
the operator $\hat{O}$.
If one chooses a non-interacting or mean-field approximated wave
function as $|\Psi^N_0 \rangle$, the expectation value in the
r.h.s.\ of Eq.~(\ref{renorm}) can be easily evaluated.
The renormalization factor for the hopping term denoted by $g^t$ is
smaller than unity because it is more difficult to hop in the presence
of the strong on-site Coulomb repulsion between electrons; that for the
exchange interaction denoted by $g^s$ is larger than unity because each
site is more often singly occupied to avoid the other electrons.
The GA was first introduced for the Hubbard model by Gutzwiller
\cite{Gutzwiller64,Gutzwiller65}, then reformulated by Ogawa {\it et
al.}\cite{Ogawa75}  A clear description of the method has been given by
Vollhardt\cite{Vollhardt84}. It was also applied to
a mean-field theory for the $t$-$J$ model by Zhang {\it et al.}\cite{FCZhang88}
Improvements of the GA by taking more intersite correlations have been
made by several authors.\cite{TCHsu90,Sigrist94,Ogata03}
The GA usually produces qualitatively correct results
although it is reported that there are also qualitative differences
in some cases\cite{CPChou06}.

The original formulation of the GA implicitly assumes that a
wave function before the projection has a fixed particle
number $N$ (in the following, we call it the ``canonical scheme'').
If the particle number of a wave function has fluctuation
(the``grand canonical scheme''), then the Gutzwiller projection reduces
the particle number (see Appendix \ref{sec:diffscheme}).
Such reduction of the particle number may arouse a question whether the
GA as Eq.~(\ref{renorm}) is valid because this equation seems to say
that the wave functions before and after the projection have similar
properties except for the double occupancy; are they similar if
they have different particle numbers?
To avoid such an unclear path,
Anderson and Ong\cite{Anderson06},
and Edegger {\it et al.}\cite{Edegger05}\ formulated a grand canonical GA by
taking the canonical scheme as a guide.
Namely, one can force the projection not to change the {\it average}
particle number, by gluing to $P_{\rm G}$ a fugacity factor that
compensates the particle-number reduction.
To our knowledge, the fugacity factor was first seen in a preliminary
form in the paper by Yokoyama and Shiba\cite{Yokoyama88} to relate the
canonical and the grand canonical VMC.  Gebhard\cite{Gebhard90}
introduced position- and spin-dependent fugacity factors for
calculational convenience of the $1/d$ expansion
whose $d\rightarrow \infty$ limit corresponds to the GA.
They also appear in the construction of the gossamer superconductivity by
Laughlin\cite{Laughlin02}.
Then, Wang {\it et al.}\cite{QHWang06}\ used position-dependent but
spin-independent fugacity factors for inhomogeneous systems.

The fugacity factors allow us freedom to choose a relation between
the particle numbers before and after the projection, and the
renormalization depends on this choice.
Recently, Ko {\it et al.}\cite{WHKo07}\ pointed out that two
contradictory formulae of the Gutzwiller renormalization factors in the
literature actually come from two different choices of the fugacity
factors.  That is, (i) the fugacity factors are determined so that the
projection conserves the {\it local} particle density of each spin direction
at each site, or (ii) so that the projection conserves the {\it total} particle
number for each spin direction (this is the usual canonical-scheme
constraint).
Mainly for the square lattice antiferromagnet, they used the canonical
scheme, and introduced additional position- and spin-dependent fugacity
factors, then calculated each renormalization factor as a ratio of
probabilities for the physical process.

In this paper, we examine in detail several different choices of
fugacity factors that impose local/global spin-dependent/independent
particle-number conservation.
We adopt the grand canonical scheme, and derive general formulae.  Some
of our formulae are different from those by the canonical derivation.
Furthermore, corrections to the conventional GA are also estimated or
derived by taking intersite correlations into account.
The structure of the paper is as follows:
Secs.~\ref{sec:local} and \ref{sec:localfug-xstates} are devoted for the
case (i), and Sec.~\ref{sec:global} for (ii).
First in Sec.~\ref{sec:local}, we derive renormalization of the hopping and
the pairing amplitude, the local spin moments and the exchange
interaction from the local spin-dependent constraint.  We test the
formulae of the hopping amplitude by comparing with the VMC.  Physical
interpretations are given for newly derived terms.
Subsequently in Sec.~\ref{sec:localfug-xstates}, we also check
orthogonality and excitation energies of projected Bogoljubov
quasiparticle states, and discuss asymmetry between positive and
negative bias spectra.
Next, in Sec.~\ref{sec:global}, formulae from the global spin-dependent
constraint are derived.  The formulation there includes cases where the
particle numbers before and after the projection are unequal.
In addition, grand canonical GAs with local/global spin-{\it independent}
constraints are briefly discussed in Sec.~\ref{sec:cano-gcano}.

In our impression, the grand canonical scheme simplifies calculation in
many cases because it is free from complicated configuration counting.
Furthermore, systematic improvement is straightforward by including
terms from larger clusters in the linked-cluster
expansion.\cite{Gutzwiller63}
The formulation we use is similar to the $1/d$ expansion by Metzner and
Vollhardt\cite{Metzner88}, and Gebhard\cite{Gebhard90}.  The
lowest-order theory in the uniform non-superconducting limit of our
formulation for the case (i) is equivalent to $d\rightarrow \infty$
limit of the $1/d$ expansion.
However, in inhomogeneous systems and in the presence of the second and
the third neighbor hopping, it is not clear if $1/d$ is a good expansion
parameter.  In addition, considering future improvements of the theory,
it may be difficult to define terms of very high order in $1/d$.
Therefore, we naively use the linked-cluster expansion as
Gutzwiller's original formulation\cite{Gutzwiller63}, then expand it
in a power series of intersite contractions and neglect high order
terms.  Furthermore, we do not adhere to making derived formulae into
the form of Eq.~(\ref{renorm}).

Throughout this paper, we use the following notation:  A wave function
before a projection is denoted by $|\Psi_0\rangle$ and it does not have
a definite particle number and may have some inhomogeneity in general.
Then, the wave function after the projection is represented by
$|\Psi\rangle = P |\Psi_0\rangle$, where $P$
is a generalized projector that includes fugacity factors defined later.
The expectation values of an arbitrary operator $\hat{O}$ by these wave
functions are denoted by
\begin{equation}
\langle \hat{O} \rangle \equiv \frac {\langle \Psi|
 \hat{O}|\Psi\rangle}{\langle \Psi |\Psi\rangle}
~,
\qquad
\langle \hat{O} \rangle_0 \equiv
\frac{\langle \Psi_0| \hat{O}|\Psi_0\rangle}
{\langle \Psi_0|\Psi_0\rangle}
~.
\end{equation}
Furthermore,
\begin{gather}
 n_{i\sigma} \equiv \langle \hat{n}_{i\sigma} \rangle_0,~
n_{ij\sigma} \equiv \langle c^\dagger_{i\sigma}c_{j\sigma} \rangle_0,~
\Delta_{ij} \equiv
\langle c_{j\downarrow}c_{i\uparrow} \rangle_0
,
\\
n_i \equiv n_{i\uparrow} + n_{i\downarrow},
 \quad
m_i \equiv \frac12 (n_{i\uparrow} - n_{i\downarrow})
.
\end{gather}
In addition, ${\bf S}_i$ denotes the spin operator at site $i$.

\section{Local constraint}
\label{sec:local}

The Gutzwiller projection changes electron-density distribution in
inhomogeneous systems in general.  However, by introducing fugacity
factors, one can force desired electron-density distribution.
We prefer to start from the grand canonical GA with a local 
constraint for each spin direction, namely,
\begin{equation}
\langle \hat{n}_{i\sigma} \rangle = \langle \hat{n}_{i\sigma} \rangle_0
~,
\label{eq:localconst}
\end{equation}
for any $i$ and $\sigma$.  Note that this local constraint is different
from the canonical scheme constraint that conserves the {\it total}
particle number.  However, this ``local canonical'' constraint
simplifies the resultant formulae as shown in the following.  For
example, some of low order corrections to the GA vanish automatically.
Furthermore, with this constraint, projected Bogoljubov quasiparticle
states are approximately orthogonal to each other, and excitation energies are
approximatively obtained by diagonalizing a renormalized Hamiltonian
(shown in Sec.~\ref{sec:localfug-xstates}).

In general, $\langle S^x_i\rangle_0$ and $\langle S^y_i\rangle_0$ can be
finite.  Such cases will be discussed only in Sec.~\ref{sec:finiteSxy},
and otherwise $\langle S^x_i \rangle_0 = \langle S^y_i \rangle_0 =0$ and
$\langle c^\dagger_{i\sigma} c_{j\bar{\sigma}} \rangle_0 = 0$ are
assumed.
Furthermore, although we have $d$-wave superconductors in mind,
there may be deviation from $d$-wave in inhomogeneous magnetic systems,
and $\langle c_{i\uparrow}^\dagger c_{i\downarrow}^\dagger \rangle_0$
(on-site pairing {\it before} the projection)
can be non-zero.
We discuss effect of $\langle c_{i\uparrow}^\dagger
c_{i\downarrow}^\dagger \rangle_0\neq 0$ in Sec.~\ref{sec:swavedos},
and otherwise assume $\langle c_{i\uparrow}^\dagger
c_{i\downarrow}^\dagger \rangle_0 =0$ for any $i$.
We also do not consider triplet pairing of the form $\langle
c_{i\sigma}^\dagger c_{j\sigma}^\dagger \rangle_0$ and set it to zero
for any $i$, $j$, $\sigma$.  The generalization to $\langle
c_{i\sigma}^\dagger c_{j\sigma}^\dagger \rangle_0 \neq 0$ is
straightforward.

\subsection{Condition for fugacity factors}

The projected wave function is defined as $|\Psi\rangle = P |\Psi_0
\rangle$ with $P \equiv  \prod_i P_i$, where
\begin{equation}
\quad P_i \equiv
 \lambda_{i\uparrow}^{\frac12\hat{n}_{i\uparrow} }
 \lambda_{i\downarrow}^{\frac12 \hat{n}_{i\downarrow}}
 (1-\hat{n}_{i\uparrow} \hat{n}_{i\downarrow})
~.
\end{equation}
The local up and down particle numbers are controlled by
$\lambda_{i\sigma}^{\frac12\hat{n}_{i\sigma}}$,
and the fugacity factors $\lambda_{i\sigma}$ will be
determined later to satisfy Eq.~(\ref{eq:localconst}).
In order to derive their explicit forms,
let us calculate the density of $\sigma$-spin electron at site $i$,
\begin{equation}
\langle  \hat{n}_{i\sigma} \rangle
=
\frac{
\left\langle 
\lambda_{i\sigma} \hat{n}_{i\sigma} ( 1 - \hat{n}_{i\bar\sigma} )
 \prod_{l\ne i} P_l^2
\right\rangle_0}
{
\left\langle
 \prod_l P_l^2
\right\rangle_0}
~.
\label{eq:niup}
\end{equation}
In principle, by applying the Wick theorem, these expectation values can
be exactly evaluated.  In practice, however, such calculation is quite
difficult to carry out because too many terms appear by the Wick
decomposition.  To approximate it, remember that intersite contractions,
$n_{ij\sigma}$ and $\Delta_{ij}$, are much smaller than on-site
contractions, $n_{i\sigma}$.  An approximation to take the leading order
with respect to the intersite contractions corresponds to the GA.  Here,
we take only on-site contractions.  Then, $l\neq i$ terms cancel out
between the numerator and the denominator, namely,
\begin{eqnarray}
 \langle \hat{n}_{i\sigma} \rangle
& \approx &
\frac
{
\lambda_{i\sigma}
( 1 - {n}_{i\bar{\sigma}} )
}
{\Xi_i} \, {n}_{i\sigma}
~,
\label{eq:nrenormlocal}
\\
 \Xi_i \equiv \langle P_i^2 \rangle_0 &=&
(1-{n}_{i\uparrow})(1-{n}_{i\downarrow})
+ \lambda_{i\uparrow} {n}_{i\uparrow}(1-{n}_{i\downarrow})
\nonumber \\&&
+ \lambda_{i\downarrow} {n}_{i\downarrow} (1-{n}_{i\uparrow})
.
\end{eqnarray}
Therefore, the condition to determine $\lambda_{i\sigma}$ is given by
$
\lambda_{i\sigma}
( 1 - {n}_{i\bar{\sigma}} )
/{\Xi_i}
 = 1
$.
By solving the simultaneous equations for up and down spins, we obtain
\begin{equation}
\lambda_{i\sigma} \approx \frac{1-n_{i\sigma}}
{ 1-n_{i} }
~, \quad
\Xi_i \approx  \frac{ (1-n_{i\uparrow}) (1-n_{i\downarrow}) }
{ 1-n_{i} }
~.
\label{eq:deflambdai}
\end{equation}

The corrections to $\langle P^2 \rangle_0 $ and $\langle
n_{i\sigma}\rangle $ can be calculated by taking into account intersite
contractions between site $i$ and other sites $l\neq i$.
Let us calculate terms proportional to $|n_{il\uparrow}|^2$.  Such terms
appear by the Wick decomposition of $\langle \hat{n}_{i\uparrow}
P^2\rangle_0$.  We take on-site contractions for the sites other than
$i,l$, and thus we only need to consider $\langle \hat{n}_{i\uparrow}
P_l^2\rangle_0$.
The operators in $\hat{n}_{i\uparrow}=c_{i\uparrow}^\dagger
c_{i\uparrow}$ are contracted with those in $c_{l\uparrow}^\dagger
c_{l\uparrow}$ or $c_{l\uparrow} c_{l\uparrow}^\dagger$ in $P_l^2$.
Then, the operators for the down spin are replaced by $n_{l\downarrow}$ or
$1-n_{l\downarrow}$.  Namely, such contribution is written as
\[
 |n_{il\uparrow}|^2
\bigg(
(1-n_{l\downarrow})
-\lambda_{l\uparrow}(1-n_{l\downarrow})
+\lambda_{l\downarrow} n_{l\downarrow}
\bigg)=0
~.
\]
In other words, the terms proportional to $|n_{il\uparrow}|^2$ vanish when
$\lambda_{i\sigma}$ is set as Eq.~(\ref{eq:deflambdai}).
Similarly, terms proportional to $\Delta_{il}^2$ also vanish.
Therefore, with Eq.~(\ref{eq:deflambdai}), we have $ \langle
\hat{n}_{i\sigma} \rangle = n_{i\sigma} + O(n_{ij\sigma}^4) +
O(\Delta_{ij}^4) $.  Estimated corrections to
$\lambda_{i\sigma}$ are also of the order of $n_{ij\sigma}^4$ or
$\Delta_{ij}^4$.

\subsection{Hopping and pairing amplitude}

For the hopping term, similar calculation can be carried out. Namely,
for $i\neq j$,
\begin{eqnarray}
\langle
c_{i\uparrow}^\dagger c_{j\uparrow}
\rangle
 & = &
\lambda_{i\uparrow}^\frac12 \lambda_{j\uparrow}^\frac12
 \frac{\displaystyle
\left\langle
c_{i\uparrow}^\dagger c_{i\downarrow} c_{i\downarrow}^\dagger
c_{j\uparrow} c_{j\downarrow} c_{j\downarrow}^\dagger
 \prod_{l\neq i,j}
P_l^2
\right\rangle_0}
{\langle P^2 \rangle_0}
\label{eq:cccccc}
\\
 & \approx &
\frac{ \lambda_{i\uparrow}^\frac12 \lambda_{j\uparrow}^\frac12
(1-n_{i\downarrow}) (1-n_{j\downarrow}) }
{ \Xi_i \Xi_j }
\langle
c_{i\uparrow}^\dagger c_{j\uparrow}
\rangle_0
~,
\end{eqnarray}
where we took on-site contractions except one intersite contraction
(that is necessary)
in the numerator.  Then, the Gutzwiller renormalization factor is given
by\cite{WHKo07}
\begin{equation}
\frac{\langle
c_{i\sigma}^\dagger c_{j\sigma}
\rangle}
{\langle
c_{i\sigma}^\dagger c_{j\sigma}
\rangle_0}
\approx
 \sqrt{ \frac{ 1-n_{i} }{1-n_{i\sigma}} }
  \sqrt{ \frac{ 1-n_{j} }{1-n_{j\sigma}} }
\equiv
g_{ij\sigma}^{t0}
~.
\label{eq:gti}
\end{equation}

\begin{figure}[h]
\includegraphics[width=6cm]{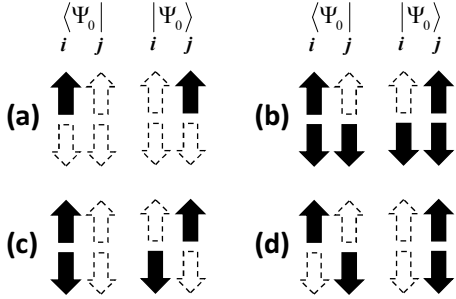} \caption{ Configurations
contributing to $\langle c_{i\uparrow}^\dagger c_{j\uparrow} \rangle_0$.
Filled arrows represent occupied states, and open dashed arrows
represent unoccupied states. Only (a) contributes to 
$\langle c_{i\uparrow}^\dagger c_{j\uparrow} \rangle$.
\label{fig:spin1}}
\end{figure}

The next order in fact involves one more site other than $i$ and $j$, but
the second and the third order of the intersite contractions for such
contribution vanish when $\lambda_{i\sigma}$ is set as
Eq.~(\ref{eq:deflambdai}).  Therefore, the third order term involves
only sites $i$ and $j$.  Namely, taking more contractions between $i$ and
$j$ in Eq.~(\ref{eq:cccccc}),
\begin{equation}
\langle
c_{i\uparrow}^\dagger c_{j\uparrow}
\rangle
  \approx
g_{ij\uparrow}^{t0}\left( n_{ij\uparrow}
-
 n_{ij\downarrow}
\frac{
n_{ij\uparrow} n_{ij\downarrow}^* + \Delta_{ij}^* \Delta_{ji}
}
{(1-n_{i\downarrow})(1-n_{j\downarrow})}
\right)
~.
\label{eq:generalizedgt}
\end{equation}
The formula for $\langle c_{i\downarrow}^\dagger c_{j\downarrow}\rangle$
is obtained by replacing as $\uparrow \Leftrightarrow \downarrow$ and
$\Delta_{ij} \Rightarrow - \Delta_{ji}$.
The $n_{ij\uparrow}|n_{ij\downarrow}|^2$ term in
Eq.~(\ref{eq:generalizedgt}) is from repulsive correlation between
down-spin holes due to the Pauli principle: all of the
four configurations in Fig.~\ref{fig:spin1} contribute to $\langle
c_{i\uparrow}^\dagger c_{j\uparrow} \rangle_0$, but only (a) does to
$\langle c_{i\uparrow}^\dagger c_{j\uparrow} \rangle$.  Then, taking into
account repulsion between down-spin holes, (a) has less weight than the
estimate by the conventional GA that neglects this correlation.

\begin{figure}[h]
\includegraphics[width=8cm]{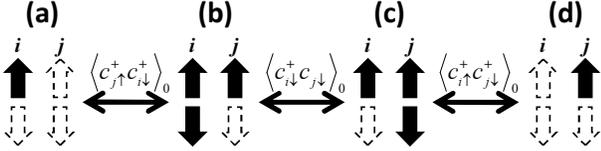} \caption{ Roundabout correlation
between (a) and (d) via (b) and (c).  These configurations in
$|\Psi_0\rangle$ correlate in the presence of superconductivity.
\label{fig:spin2}}
\end{figure}

On the other hand, the $n_{ij\downarrow} \Delta_{ij}^* \Delta_{ji}$ term
is from superconducting correlation; negative for
$\Delta_{ij}=\Delta_{ji}$ (singlet), and positive for
$\Delta_{ij}=-\Delta_{ji}$ (triplet).
This term seems related to the phase difference between the four
configurations in Fig.~\ref{fig:spin2}, which appear in $|\Psi\rangle_0$
(before the projection).  Our rough explanation 
in the case of $n_{ij\uparrow} \simeq n_{ij\downarrow} $ is as follows:
Suppose $\zeta_a$, $\zeta_b$, $\zeta_c$, $\zeta_d$ are coefficients of
the configuration (a,b,c,d) in $|\Psi\rangle_0$, and assume they are
real numbers.
Then, $\zeta_a \zeta_d$ contributes to $n_{ij\uparrow}$, and $-\zeta_b
\zeta_c$ contributes to $n_{ij\downarrow}$.
Remember that the conventional GA can be derived by taking the ratio of
the probability of configurations\cite{FCZhang88,WHKo07}; it implicitly
assumes that $\zeta_a \zeta_d$ and $-\zeta_b \zeta_c$ have the same sign.
Turning on the superconducting correlation, configurations (a) and (b)
as well as (c) and (d) start to correlate.
Then, their contribution to the singlet order parameter before the
projection $\langle c_{i\uparrow}^\dagger c_{j\downarrow}^\dagger -
c_{i\downarrow}^\dagger c_{j\uparrow}^\dagger \rangle_0$ is proportional
to $ \zeta_b \zeta_a + \zeta_c \zeta_d $.
The magnitude of this quantity, however, is small if $\zeta_a \zeta_d$
and $-\zeta_b \zeta_c$ have the same sign.
Therefore, to strengthen the singlet superconducting correlation,
all of $\zeta_a \zeta_d,\zeta_b \zeta_c$ should be small. Accordingly,
the weight of Fig.~\ref{fig:spin1}(a) should be smaller than the
estimate by the conventional GA.

\begin{figure}[h]
\includegraphics[width=8cm]{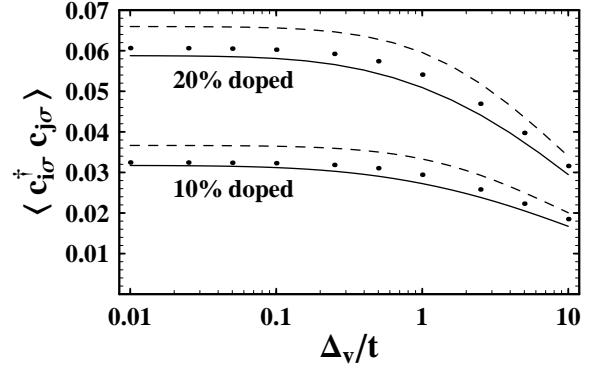} \caption{The neatest-neighbor
hopping amplitude $\langle c_{i\sigma}^\dagger c_{j\sigma} \rangle$
calculated by the conventional GA [Eq.~(\ref{eq:gti}), broken line], the
generalized GA [Eq.~(\ref{eq:generalizedgt}), solid line], and the VMC
[dots] for the projected uniform non-magnetic BCS $d$-wave
superconductor.  \label{fig:hopping}}
\end{figure}

We test this formula for a simplest case, {\it i.e.},
the standard uniform non-magnetic BCS $d$-wave superconductor,
\begin{gather}
 |\Psi_0 \rangle = \prod_k
( u_k + v_k c_{k,\uparrow}^\dagger c_{k,\downarrow}^\dagger )|0\rangle
~,
\label{eq:dwaveBCS}
\\
u_k \equiv \sqrt{ \frac{1}{2} \left( 1 + \frac{\xi_k}{E_k}  \right)}~,
\quad
v_k \equiv \frac{\Delta_k}{|\Delta_k|}
\sqrt{ \frac{1}{2} \left( 1 - \frac{\xi_k}{E_k}  \right)} ~,
\nonumber\\
E_k \equiv \sqrt{ \xi_k^2 + \Delta_k^2 } ~, \quad
\Delta_k \equiv \Delta_{\rm v}\,(\cos k_x -\cos k_y ) ~, \nonumber \\
\xi_k \equiv -2 t\,(\cos k_x +\cos k_y) -\mu~. \nonumber
\end{gather}
The conventional GA as Eq.~(\ref{eq:gti}), the generalized GA as
Eq.~(\ref{eq:generalizedgt}), and the VMC are compared in
Fig.~\ref{fig:hopping} for the nearest-neighbor hopping.
Here, the generalized GA is done using 200$\times$200 sites, and
practically the finite-size effects are negligible;
errors only come from neglect of the higher order of the intersite
contractions.
$\mu$ is adjusted to satisfy each hole concentration for each point. 
The VMC is carried out using 30$\times$30 sites with $x$-antiperiodic
$y$-periodic boundary condition.  The hopping amplitude is averaged
over every bond, and the statistical errors are negligible in the scale
of this figure.  For comparison with the GAs, $\mu$ is also
adjusted to equalize the doping before the projection with that after.
At small $\Delta_{\rm v}$,
the generalized GA agrees with the VMC very well.
As $\Delta_{\rm v}$ increases, the deviation becomes larger.
This is possibly because $O(\Delta_{ij}^4)$ term neglected in
Eq.~(\ref{eq:nrenormlocal}) may start to make an important contribution.

\begin{figure}[h]
\includegraphics[width=8cm]{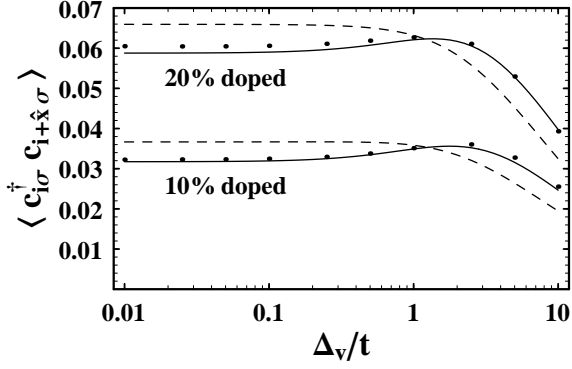}
\caption{The neatest-neighbor hopping amplitude of
the projected ``$p$-wave'' superconductor
calculated by the conventional GA [Eq.~(\ref{eq:gti}), broken line], the
generalized GA [Eq.~(\ref{eq:generalizedgt}), solid line], and the VMC
[dots].
\label{fig:hoppingp}}
\end{figure}

In inhomogeneous systems, there may be deviation from the $d$-wave.
Since it is rather difficult to force the local constraint
for the VMC in inhomogeneous systems,
let us test the formula using a simpler non-$d$-wave, namely, 
a uniform $p$-wave superconductor by redefining
\begin{equation}
\Delta_k \equiv \Delta_{\rm v} \sin k_x
\end{equation}
in Eq.~(\ref{eq:dwaveBCS}).  The nearest-neighbor hopping amplitude in
the $x$-direction is plotted in Fig.~\ref{fig:hoppingp}.  The
generalized GA shows a good overall agreement with the VMC.  It
especially reproduces characteristic peak at $\Delta_{\rm v}\sim 2t$
caused by the $n_{ij\downarrow} \Delta_{ij}^* \Delta_{ji}$ term in
contrast to the conventional GA.

\begin{figure}[h]
\includegraphics[width=8cm]{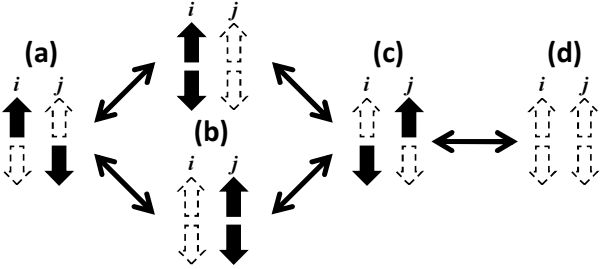} \caption{ Roundabout correlation
between (a) and (d) via (b) and (c) in $|\Psi_0\rangle$.
\label{fig:spinDelta}}
\end{figure}

The superconducting order parameters
$\langle c_{i\uparrow}^\dagger c_{j\downarrow}^\dagger \rangle$
can be calculated similarly to the hopping term, {\it i.e.},
\begin{equation}
\langle c_{i\uparrow}^\dagger c_{j\downarrow}^\dagger \rangle
= \sqrt{g_{ii\uparrow} \, g_{jj\downarrow}}
 \left( \Delta_{ij}^* 
+
 \Delta_{ji}^*
\frac{
n_{ij\uparrow} n_{ij\downarrow}^* + \Delta_{ij}^* \Delta_{ji}
}
{(1-n_{i\downarrow})(1-n_{j\uparrow})}
\right)
.
\label{eq:Deltarenorm}
\end{equation}
The $\Delta_{ij}^*$ term represents the direct correlation between the
$i\uparrow, j\downarrow$ occupied state [Fig.~\ref{fig:spinDelta}(a)]
and the empty state [Fig.~\ref{fig:spinDelta}(d)].  The $\Delta_{ij}^*
|\Delta_{ji}|^2$ term contains the attractive correlation between holes
of $i\downarrow$ and $j\uparrow$; if $\Delta_{ji}$ is finite,
$i\downarrow$ and $j\uparrow$ tend to be simultaneously occupied or
unoccupied, and it is less likely that only one of them is occupied.
Accordingly, this effect increases weight of the configurations in
Fig.~\ref{fig:spinDelta}(a) and (d), and appears as the positive
correction in Eq.~(\ref{eq:Deltarenorm}).
The $\Delta_{ij}^* n_{ij\uparrow} n_{ij\downarrow}^*$ term represents
roundabout correlation between $i\uparrow$ and $j\downarrow$ through
$i\downarrow$ and $j\uparrow$ as depicted in Fig.~\ref{fig:spinDelta}.
Argument similar to what is used for the hopping amplitude
(Fig.~\ref{fig:spin2}) leads to the conclusion that the singlet
correlation enhances weight of configurations in
Fig.~\ref{fig:spinDelta} in this case.

Note that Eqs.~(\ref{eq:generalizedgt},\ref{eq:Deltarenorm}) is mainly
aimed at $|i-j|=1$.  For next-nearest neighbors, $O(n_{ij}^4)$ of
$|i-j|=1$ may be comparable to $O(n_{i'j'}^2)$ of $|i'-j'|=2$ and the
former may be dominant especially in high dimensions.  In general, as
$i$ and $j$ separate from each other, the approximation by
Eqs.~(\ref{eq:generalizedgt},\ref{eq:Deltarenorm}) may lose accuracy.

\subsection{Spin moment and exchange interaction}
\label{sec:localex}

By definition, the local spin-$z$ component at each site is not
renormalized, {\it i.e.},
\begin{equation}
  \langle S^z_i \rangle
=
  \langle S^z_i \rangle_0 = m_i
~.
\end{equation}
For the exchange interaction term $\langle {\bf S}_{i} \cdot {\bf S}_{j}
\rangle $, we take up to the second order of intersite contractions.
Using symbols $\uparrow,\downarrow$, and $+1,-1$, interchangeably,
it is written as
\begin{eqnarray}
&& \langle S^z_{i} S^z_{j}\rangle
 =
\frac1{4\langle P^2 \rangle_0}
\sum_{\sigma,\tau=\pm1}\sigma \tau
\lambda_{i\sigma}\lambda_{j\tau}
\nonumber \\&&  \qquad\qquad\qquad
\Big \langle
c_{i\sigma}^\dagger c_{i\sigma} c_{i\bar\sigma} c_{i\bar\sigma}^\dagger
c_{j\tau}^\dagger c_{j\tau} c_{j\bar\tau} c_{j\bar\tau}^\dagger
\prod_{l\neq i,j}
P_l^2
\Big \rangle_0
\nonumber \\&&
 \approx
 m_{i} m_{j}-\frac14
\Big[
(1-n_{i\uparrow})(1-n_{i\downarrow})
(1-n_{j\uparrow})(1-n_{j\downarrow})
\Big]^{-1}
\nonumber \\&& \quad
\times \Big[
|n_{ij\uparrow}|^2 (1-2m_i)(1-2m_j)(1-n_{i\downarrow})(1-n_{j\downarrow})
\nonumber \\&& \quad
+|n_{ij\downarrow}|^2 (1+2m_i)(1+2m_j) (1-n_{i\uparrow})(1-n_{j\uparrow})
\nonumber \\&& \quad
+|\Delta_{ij}|^2 (1-2m_i)(1+2m_j) (1-n_{i\downarrow})(1-n_{j\uparrow})
\nonumber \\&& \quad
+|\Delta_{ji}|^2 (1+2m_i)(1-2m_j) (1-n_{i\uparrow})(1-n_{j\downarrow})
\Big],
\label{eq:szszrenorm}
\end{eqnarray}
\begin{eqnarray}
&&
 \langle
 S^x_{i} S^x_{j}+S^y_{i} S^y_{j}
\rangle
\nonumber \\
&& \qquad
 =
\frac{
\sqrt{\lambda_{i\uparrow}\lambda_{i\downarrow}
 \lambda_{j\uparrow} \lambda_{j\downarrow}}
}{2\langle P^2 \rangle_0}
\sum_{\sigma}
\Big \langle
c_{i\sigma}^\dagger c_{i\bar\sigma} c_{j\bar\sigma}^\dagger c_{j \sigma}
\prod_{l\neq i,j}
P_l^2
\Big \rangle_0
\nonumber \\
&& \qquad \approx
\frac{-\,{\rm Re}\big[
 n_{ij\uparrow}n_{ij\downarrow}^* + \Delta_{ij} \Delta_{ji}^* \big]}
{\sqrt{
(1-n_{i\uparrow})(1-n_{i\downarrow})
(1-n_{j\uparrow})(1-n_{j\downarrow})
}}
\nonumber \\
& & \qquad =
g^{sxy}_{ij}
\langle
S^x_{i} S^x_{j}+S^y_{i} S^y_{j}
\rangle_0
~,
\label{eq:sxsxrenorm}
\end{eqnarray}
with
\begin{equation}
g^{sxy}_{ij} \equiv
\frac{1}{\sqrt{(1-n_{i\uparrow})(1-n_{i\downarrow})
(1-n_{j\uparrow})(1-n_{j\downarrow})}}
~.
\end{equation}
Here, Eq.~(\ref{eq:szszrenorm}) seems different from what is derived as
a ratio of probabilities for the physical process using the canonical
scheme with the fugacity factors by Ko {\it et al.}\cite{WHKo07} We
speculate that it possibly does not take into account all of the
contractions above.

To compare with the result by $1/d$ expansion by
Gebhard\cite{Gebhard90}, set $\Delta_{ij}=\Delta_{ji}=0$ and consider
antiferromagnets. By setting $n_{i\sigma}=n_{j\bar\sigma}$ in
Eqs.~(\ref{eq:szszrenorm},\ref{eq:sxsxrenorm}), these equations are
reduced to
\begin{gather}
 \langle S^z_{i} S^z_{j}\rangle
 \approx
 m_{i} m_{j}-
\frac{ (1-4m_i^2)
\Big(  \langle S^z_{i} S^z_{j}\rangle_0 -m_i m_j \Big)
 }{ (1-n_{i\uparrow})(1-n_{i\downarrow}) }
\label{eq:1dxgsz}
\\
 \langle S^x_{i} S^x_{j}+S^y_{i} S^y_{j} \rangle
 \approx
\frac{
\langle
S^x_{i} S^x_{j}+S^y_{i} S^y_{j}
\rangle_0
} { (1-n_{i\uparrow})(1-n_{i\downarrow}) }
~,
\label{eq:1dxgsxy}
\end{gather}
which are equivalent to the formula by the $1/d$ expansion\cite{Gebhard90}.
However, when $\Delta_{ij}\neq 0$, renormalization of $ \langle S^z_{i}
S^z_{j}\rangle$ is not reduced to such a simple form, and we need the
original formula, Eq.~(\ref{eq:szszrenorm}).
Note that Eqs.~(\ref{eq:1dxgsz},\ref{eq:1dxgsxy})
can be used also for non-superconducting ferromagnets.
Namely, the local constraint leads to the conclusion that
antiferromagnets and ferromagnets are renormalized similarly.
This is in distinct contrast to results of the GA with {\it global}
constraint as will be discussed in Sec.~\ref{sec:ferro}.

\subsection{Systems with nonzero spin-$xy$ components}
\label{sec:finiteSxy}

This choice of fugacity factors encounters difficulties when $\langle
S^x_i \rangle_0$ or $\langle S^y_i\rangle_0$ is finite.  Let us redo the
derivation including ${\cal S}^{\pm}_{i} \equiv \langle S^{\pm}_{i}
\rangle_0$:
\begin{eqnarray}
 \langle \hat{n}_{i\sigma} \rangle
 \approx 
\frac
{
\lambda_{i\sigma} [
{n}_{i\sigma} ( 1 - {n}_{i\bar\sigma} )
+{\cal S}^+_i{\cal S}^-_i
]
}
{\Xi_i}
~,
\label{eq:nrenormSS}
\end{eqnarray}
\begin{eqnarray}
&& \Xi_i =
(1-{n}_{i\uparrow})(1-{n}_{i\downarrow})
+ \lambda_{i\uparrow} {n}_{i\uparrow}(1-{n}_{i\downarrow})
\nonumber \\&& \qquad
+ \lambda_{i\downarrow} {n}_{i\downarrow} (1-{n}_{i\uparrow})
+(\lambda_{i\uparrow}+\lambda_{i\downarrow}-1){\cal S}^+_i{\cal S}^-_i
.
\end{eqnarray}
The condition to determine $\lambda_{i\sigma}$ is
\begin{equation}
\lambda_{i\sigma}
\frac
{
{n}_{i\sigma} ( 1 - {n}_{i\bar{\sigma}} )
+ {\cal S}^+_i{\cal S}^-_i
}
{\Xi_i}  = n_{i\sigma}
~.
\label{eq:forlambdaSS}
\end{equation}
This is solved to give
\begin{eqnarray}
 \Xi_i  & \approx &
\frac{(1-n_{i\uparrow}) (1-n_{i\downarrow}) - {\cal S}^+_i{\cal S}^-_i}
{1-n_i}
~, \\
\lambda_{i\sigma} & \approx &
\frac{n_{i\sigma}}{n_{i\sigma}(1-n_{i\bar\sigma})+{\cal S}^+_i{\cal S}^-_i}
~ \Xi_i
~.
\end{eqnarray}

For a spin moment,
$
 \langle  S^z_{i} \rangle
= \langle  S^z_{i} \rangle_0
$, and
\begin{multline}
\langle
S^\pm_{i}
\rangle
  \approx 
\frac{\sqrt{\lambda_{i\uparrow}\lambda_{i\downarrow}}}{\Xi_i}
{\cal S}^\pm_i
\\
 = 
{\cal S}^\pm_i
\sqrt{
\frac{n_{i\uparrow}}{n_{i\uparrow}(1-n_{i\downarrow})
+|{\cal S}^+_i|^2
}
} 
\sqrt{
\frac{n_{i\downarrow}}{n_{i\downarrow}{(1-n_{i\uparrow})}
+|{\cal S}^+_i|^2
 }
}
~.
\end{multline}
This renormalization factor for ${\cal S}^\pm_i$ is larger than unity
because it is not bound by the local constraint.
Since $xy$ component is renormalized differently from $z$ component,
approximation depends on humans' choice of $z$-axis.  This asymmetry is
probably related to what is discussed by Ko {\it et al.}\cite{WHKo07}
The most reasonable choice of $z$-axis we think is making it parallel to
$\langle {\bf S}_i \rangle_0$ at each site.  Then, ${\cal S}^\pm_i=0$
for any $i$.  It is equivalent to formulating a GA with constraints
$\langle \hat{n}_{i\uparrow}+\hat{n}_{i\downarrow} \rangle = n_i$ and
$\langle {\bf S}_i \rangle = \langle {\bf S}_i \rangle_0$.  However,
such a GA may yield very complicated renormalization factors for
intersite terms.  One way to avoid such a complexity is to use
spin-independent constraint as shown in Sec.~\ref{sec:cano-gcanolocal}.

\section{Local constraint: Excited states}
\label{sec:localfug-xstates}

The GA with the position- and spin-dependent constraint discussed in the
previous section has an advantage in constructing plausible excited
states which are approximately orthogonal to each other as shown below.

For shorthand notation, we use
\begin{equation}
 c_i \equiv c_{i\uparrow} ~,  \qquad
 c_{N_{\rm L}+i} \equiv c^\dagger_{i\downarrow},
\label{eq:shorthand}
\end{equation}
where $N_{\rm L}$ is the number of lattice sites.  Then, the subscript of this
new operator runs from 1 to $2N_{\rm L}$, and we represent it by single Greek
symbols as $c_\rho$.  Furthermore, we define
\begin{equation}
 \underbar{$\hat{n}$}_{\rho\zeta} \equiv c_\rho^\dagger c_\zeta ~,
\qquad
 \underbar{$n$}_{\rho\zeta} \equiv
\langle \underbar{$\hat{n}$}_{\rho\zeta} \rangle_0
~.
\end{equation}

\subsection{Bogoljubov de Gennes (BdG) equation}

As a preparation, let us begin with deriving a BdG equation by
minimizing the Gutzwiller-approximated energy following the procedure by
Wang {\it et al.}\cite{QHWang06}
In the following, we work more on general properties of a BdG equation
with the Gutzwiller projection, and do not use any Hamiltonian explicitly.
However, what we have in mind is inhomogeneous $t$-$J$--type models,
\begin{equation}
 H_{tJ}\equiv
 P_{\rm G}\left(
- \sum_{ij\sigma} t_{ij} c_{i\sigma}^\dagger c_{j\sigma}
+ \sum_{\langle i,j \rangle}J_{ij} {\bf S}_{i} \cdot {\bf S}_{j}
 \right) P_{\rm G}
         ~,
\end{equation}
where the $t_{ij}$ term with $i=j$ may represent local impurity
potentials.  The zero-temperature grand potential $ \Omega \equiv
\left\langle H_{tJ} - \mu \sum_{i,\sigma} \hat{n}_{i\sigma}
\right\rangle $ can be approximated by the GA, and represented by a
function of $\underbar{$n$}_{\rho\zeta}$, namely,
\begin{eqnarray}
 \Omega \approx
 \Omega_{\rm GA}\left[  \{\underbar{$n$}_{\rho\zeta}\},\mu \right] ~.
\end{eqnarray}
We do not show the explicit form of $\Omega_{\rm GA}$ because it can be
derived straightforwardly by using the formulae in the previous section.
In the derivation, one can choose the level of the approximation:
If one takes only the leading order of the intersite contractions,
formulae in the non-magnetic case are equivalent to those derived by
Wang {\it et al.}\cite{QHWang06}\ and Li {\it et al.}\cite{CLi06}
If an improved Gutzwiller approximation such as Eq.~(\ref{eq:generalizedgt})
is used, a more accurate solution can be obtained in principle, although
it may be more difficult to find self-consistent solutions.

The chemical potential $\mu$ is determined to adjust the particle number $N$
to satisfy
$
N= - {\partial \Omega_{\rm GA}}/{\partial \mu}
$.
The other variables are functional of $\Psi_0$ and determined by
minimizing $\Omega_{\rm GA}$,
\begin{equation}
\frac{\delta \Omega_{\rm GA}}{\delta \Psi_0} =
 \sum_{\rho\zeta}
\frac{\partial \Omega_{\rm GA}}
 {\partial  \underbar{$n$}_{\rho\zeta}}
\frac{\delta \underbar{$n$}_{\rho\zeta}}{\delta \Psi_0}
= 0
~.
\label{eq:omegaderiv}
\end{equation}
Assuming $\langle \Psi_0|\Psi_0\rangle = 1$, then
\begin{equation}
 \delta \underbar{$n$}_{\rho\zeta} =
\langle \delta \Psi_0 | \underbar{$\hat{n}$}_{\rho\zeta} |\Psi_0 \rangle
+
\langle \Psi_0 |\underbar{$\hat{n}$}_{\rho\zeta}| \delta \Psi_0 \rangle
~.
\label{eq:deltan}
\end{equation}
By combining Eqs.~(\ref{eq:omegaderiv},\ref{eq:deltan}),
\begin{gather}
 \delta \Omega_{\rm GA} = \langle \delta \Psi_0|H_{\rm BdG}|\Psi_0\rangle
+ \langle \Psi_0|H_{\rm BdG}|\delta \Psi_0\rangle
~,
\\ \text{where}\quad
H_{\rm BdG} \equiv
\sum_{\rho\zeta}
\frac{\partial \Omega_{\rm GA}}{\partial \underbar{$n$}_{\rho\zeta}}
 \underbar{$\hat{n}$}_{\rho\zeta}
~.
\end{gather}
Then, $\Omega_{\rm GA}$ takes an extremum when $|\Psi_0\rangle$ is an
eigenstate of $H_{\rm BdG}$, namely,
\begin{gather}
 H_{\rm BdG}|\Psi_0\rangle = E_{\rm BdG} |\Psi_0\rangle ~,
\label{eq:BdG1}
\\
\delta \Omega_{\rm GA} = E_{\rm BdG}\big(\,
\langle \delta \Psi_0|\Psi_0\rangle + \langle \Psi_0|\delta \Psi_0\rangle
\,\big)
= 0
~.
\end{gather}
The main differences from usual BdG Hamiltonian are the
local renormalization factors in front of $t_{ij}$ and $J_{ij}$,
and the effective local chemical potential terms $-\sum_i \mu_{i\sigma}
\hat{n}_{i\sigma}$ with
$
\mu_{i\sigma} = - {\partial \Omega_{\rm GA}}/{\partial n_{i\sigma}}
 - \mu
$
which come from $n_{i\sigma}$-dependence of the renormalization
factors.  Local modulations of $t_{ij}$ and $J_{ij}$ tend to be enhanced
by the local renormalization factors, and impurity potentials tend to be
screened by the local chemical potentials.\cite{FukushimaSNS}

\subsection{Quasi-particles}

We rewrite $H_{\rm BdG}$ in a matrix form,
\begin{equation}
 H_{\rm BdG} = \sum_{\rho,\zeta=1}^{2N_{\rm L}} c_\rho^\dagger H_{\rho\zeta}
  c_\zeta
~.
\end{equation}
The $2N_{\rm L}\times2N_{\rm L}$ matrix $H_{\rho\zeta}$ can be diagonalized using a unitary
matrix $U$, namely,
\begin{equation}
H_{\rho\zeta} = \sum_n U_{\rho n} E_n (U^\dagger)_{n \zeta}.
\end{equation}
Then, using
\begin{equation}
 \begin{aligned}
\gamma_{n} \equiv \sum_\rho (U^\dagger)_{n\rho} \, c_\rho
\qquad (E_n>0), \\
\gamma_{n}^\dagger \equiv \sum_\rho (U^\dagger)_{n\rho} \, c_\rho
\qquad (E_n<0),
\end{aligned}
\label{eq:defgamman}
\end{equation}
the Hamiltonian is diagonalized as,
\begin{equation}
H_{\rm BdG} = \sum_n E_n \gamma_{n}^\dagger \gamma_{n}.
\end{equation}
The ground state of this effective Hamiltonian is $ |\Psi_0\rangle\equiv
\prod_n \gamma_n |0\rangle$. Suppose the ground state is well
approximated by $P |\Psi_0\rangle$.  Naively, one may assume that
excited states are constructed by $ P \gamma_n^\dagger |\Psi_0\rangle $.
This form of excited states was first introduced for uniform systems by
Zhang {\it et al.}\cite{FCZhang88}
For fugacity factors in $P$, we use those in the ground state even for
the excited states.  It probably correspond to assuming that the
quasi-particles $\gamma_n$ are not very localized and that the change of
the particle distribution is negligible.

\subsection{Orthogonality of the excited states}
\label{sec:ortho}

The orthogonality of these excited states can be checked by expanding
$\gamma_n$ using Eq.~(\ref{eq:defgamman}).
For example, for $E_n>0$, $E_m>0$,
\begin{equation}
 \langle \Psi_0 | \gamma_n P \, P \gamma_m^\dagger |\Psi_0\rangle
= \sum_{\rho\zeta} U_{\rho n}^* U_{\zeta m}
 \langle c_\rho P^2 c_\zeta^\dagger \rangle_0
~.
\end{equation}
Here, we have to mind a discrepancy between creation and
annihilation operators;
\begin{gather}
P_i^2 c_{i\sigma}^\dagger =
c_{i\sigma}^\dagger 
\lambda_{i \sigma} (1-\hat{n}_{i\bar{\sigma}} )
~,
\\
P_i^2 c_{i\sigma} =
c_{i\sigma}
\big[(1-\hat{n}_{i\bar{\sigma}} ) +
\lambda_{i\bar\sigma}  \hat{n}_{i\bar{\sigma}} \big]
~.
\end{gather}
Then, as the leading-order theory, we take on-site contractions except
one intersite contraction.  Thanks to Eq.~(\ref{eq:deflambdai}),
renormalization factors are reduced to a simple form, {\it i.e.,}
\begin{equation}
\frac{ \lambda_{i \sigma} (1-n_{i\bar{\sigma}} )}{\Xi_i}
=
\frac{(1-n_{i\bar{\sigma}} ) +
\lambda_{i\bar\sigma}  n_{i\bar{\sigma}} }{\Xi_i}=1
~,
\end{equation}
and we obtain simple results,
\begin{gather}
\label{eq:cPc1}
\frac{ \langle  c_{i\sigma}^\dagger P^2 c_{j\tau} \rangle_0 }
{ \langle  P^2 \rangle_0 }
\approx
\langle c_{i\sigma}^\dagger c_{j\tau} \rangle_0
~,
\\
\frac{ \langle  c_{i\sigma} P^2 c_{j\tau}^\dagger \rangle_0 }
{ \langle  P^2 \rangle_0 }
\approx
\langle c_{i\sigma} c_{j\tau}^\dagger \rangle_0
~,
\end{gather}
\begin{gather}
\frac{ \langle  c_{i\sigma}^\dagger P^2 c_{j\tau}^\dagger \rangle_0 }
{ \langle  P^2 \rangle_0 }
\approx
\langle c_{i\sigma}^\dagger c_{j\tau}^\dagger \rangle_0
~,
\\
\frac{ \langle  c_{i\sigma} P^2 c_{j\tau} \rangle_0 }
{ \langle  P^2 \rangle_0 }
\approx
\langle c_{i\sigma} c_{j\tau} \rangle_0
~,
\label{eq:cPc4}
\end{gather}
for any $i,j,\sigma,\tau$ including $i=j$.
In general, $\langle c^\dagger_{i\sigma}P^2 c_{i\bar{\sigma}}^\dagger
\rangle_0$ does not satisfy this relation, but we are lucky enough to
use the off-site pairing assumption, $\langle c^\dagger_{i\sigma}
c_{i\bar{\sigma}}^\dagger \rangle_0 =0$, then $ \langle
c^\dagger_{i\sigma}P^2 c_{i\bar{\sigma}}^\dagger \rangle_0 \approx 0 $,
and can exclude such an exception.
Our projected superconducting state includes the Fermi sea as a special
case, and these relations look different at a sight from those derived
for the Fermi sea by Fukushima {\it et al.}\cite{Fukushima05}  In fact,
however, these are identical if one remembers that $P$ contains fugacity
factors.

Using Eq.~(\ref{eq:cPc1}-\ref{eq:cPc4}), one can transform back from
$c_\rho$ to $\gamma_n$ to yield
\begin{equation}
\frac{ \langle \Psi_0 | \gamma_n P^2 \gamma_m^\dagger |\Psi_0\rangle }
{ \langle \Psi_0 | P^2 |\Psi_0\rangle }
\approx 
\langle \gamma_n \gamma_m^\dagger \rangle_0
=\delta_{nm}
~.
\label{eq:gammagamma}
\end{equation}
That is, the excited states are orthogonal to each other within the GA.

\subsection{Excitation energy}

Let $|0\rangle$ and $|n\rangle$ denote
the {\it normalized} ground and excited states,
\begin{equation}
|0\rangle \equiv
\frac{ P |\Psi_0\rangle }
{ \sqrt{\langle \Psi_0 | P^2 |\Psi_0\rangle} }
,\quad
|n\rangle \equiv
\frac{ P \gamma_n^\dagger |\Psi_0\rangle }
{ \sqrt{\langle \Psi_0 |\gamma_n P^2 \gamma_n^\dagger|\Psi_0\rangle} }
.
\label{eq:defexcited}
\end{equation}
Neglecting the second order of the difference in
$\underbar{$n$}_{\rho\zeta}$,
\begin{eqnarray}
&& \langle n|H_{tJ}|n\rangle - \langle 0|H_{tJ}|0\rangle \nonumber \\
& \approx &
 \sum_{\rho\zeta}
\frac{\partial \Omega_{\rm GA}}{\partial  \underbar{$n$}_{\rho\zeta}}
\Big(\,
\langle n| \underbar{$n$}_{\rho\zeta} |n\rangle -
\langle 0| \underbar{$n$}_{\rho\zeta} |0\rangle
\,\Big)
\nonumber \\
&=& \langle n | H_{\rm BdG} |n\rangle - \langle 0 | H_{\rm BdG} |0\rangle
= E_n
~.
\end{eqnarray}
Therefore, the excitation energies are approximately the same as
eigenenergies of the effective Hamiltonian.

\subsection{Density of states}
To calculate the local density of states, we need matrix elements,
$|\langle n|c_{i\sigma}^\dagger|0\rangle|^2$ and
$|\langle n|c_{i\sigma}|0\rangle|^2$.
First of all, using Eq.~(\ref{eq:gammagamma}) with $n=m$, the
normalization of the excited states can be replaced as
\begin{equation}
{ \langle \Psi_0 | \gamma_n P^2 \gamma_n^\dagger |\Psi_0\rangle }
\approx 
{ \langle \Psi_0 | P^2 |\Psi_0\rangle }
~.
\end{equation}
Then, we expand $\gamma_n$ in $|n\rangle$ using Eq.~(\ref{eq:defgamman})
and use simple relations similar to Eqs.~(\ref{eq:cPc1}-\ref{eq:cPc4}),
namely,
\begin{gather}
\label{eq:cpcp1}
\frac{ \langle  c_{j\tau}^\dagger P c_{i\sigma} P \rangle_0 }
{ \langle  P^2 \rangle_0 }
\approx
\sqrt{ g_{ii\sigma}^{t0} } \,
\langle c_{j\tau}^\dagger c_{i\sigma} \rangle_0
~,
\\
\label{eq:cpcp2}
\frac{ \langle  c_{j\tau} P c_{i\sigma}^\dagger P \rangle_0 }
{ \langle  P^2 \rangle_0 }
\approx
\sqrt{ g_{ii\sigma}^{t0} } \,
\langle c_{j\tau} c_{i\sigma}^\dagger \rangle_0
~,
\end{gather}
\begin{gather}
\frac{ \langle  c_{j\tau}^\dagger P c_{i\sigma}^\dagger P \rangle_0 }
{ \langle  P^2 \rangle_0 }
\approx
\sqrt{ g_{ii\sigma}^{t0} } \,
\langle c_{j\tau}^\dagger c_{i\sigma}^\dagger \rangle_0
~,
\\
\frac{ \langle  c_{j\tau} P c_{i\sigma} P \rangle_0 }
{ \langle  P^2 \rangle_0 }
\approx
\sqrt{ g_{ii\sigma}^{t0} } \,
\langle c_{j\tau} c_{i\sigma} \rangle_0
~,
\label{eq:cpcp4}
\end{gather}
for any $i,j,\sigma,\tau$.
These formulae are true also for $i=j$, more explicitly,
\begin{gather}
\frac{ \langle  c_{i\sigma}^\dagger P c_{i\sigma} P \rangle_0 }
{ \langle  P^2 \rangle_0 }
= \frac1{\sqrt{\lambda_{i\sigma}}} \, n_{i\sigma}
~,
\label{eq:localciPciP1}
\\
\frac{ \langle  c_{i\sigma} P c_{i\sigma}^\dagger P \rangle_0 }
{ \langle  P^2 \rangle_0 }
= \sqrt{\lambda_{i\sigma}} \, (1-n_{i\uparrow}-n_{i\downarrow})
~.
\label{eq:localciPciP2}
\end{gather}
These relations are exact if exact $\lambda_{i\sigma}$ are used.

Since the indices of the renormalization factor are only from those
of the operator between two $P$'s,
we can transform back to $\gamma_n$ to yield
\begin{eqnarray}
 |\langle n|c_{\rho}^\dagger|0\rangle|^2
 &=&
 g_{ii\sigma}^{t0} \, |\langle \Psi_0 | \gamma_n 
 c_{\rho}^\dagger|\Psi_0\rangle|^2
\nonumber \\
 &=& g_{ii\sigma}^{t0} \, |U_{\rho n}|^2  \qquad (E_n >0)
~,
 \\
 |\langle n|c_{\rho}|0\rangle|^2
 &=&
 g_{ii\sigma}^{t0} \, |\langle \Psi_0 | \gamma_n 
 c_{\rho}|\Psi_0\rangle|^2
\nonumber \\
 &=& g_{ii\sigma}^{t0} \, |U_{\rho n}|^2  \qquad (E_n <0)
~,
\end{eqnarray}
where $\rho=(i,\sigma)$ as Eq.~(\ref{eq:shorthand}).  The common
renormalization factor $g_{ii\sigma}^{t0}$ tells us that the positive
and negative bias spectra are symmetric.  This symmetric density of states is
also obtained by the canonical scheme GA\cite{Fukushima05,Randeria05}.
We go one-step further about this point in the next subsection.

For $A(k,\omega)$, we need matrix elements in $k$-space, $|\langle
n|c_{k,\sigma}^\dagger|0\rangle|^2$ and $|\langle
n|c_{k,\sigma}|0\rangle|^2$, where $c_{k,\sigma}=N_{\rm L}^{-1/2}\sum_i
c_{i\sigma} \exp(i k R_i)$.  These can be obtained by the Fourier
transform of Eqs.~(\ref{eq:cpcp1}-\ref{eq:cpcp4}).

\subsection{Electron addition-removal asymmetry
caused by higher order terms
}
\label{sec:highaddrem}

The conventional BCS theory tells us that the quasi-particle excitation
spectra are symmetric between positive and negative bias.  However,
local density of states of high-$T_{\rm c}$ superconductors measured by the STM
is highly asymmetric and there is an argument that attributes this
asymmetry to strong electron correlation.\cite{Anderson06} Namely,
electron addition may be more difficult than electron removal
because the injected electron may be repelled by the other electrons due
to their strong Coulomb repulsion.
It is controversial whether the projected quasi-particle states have
symmetric spectra or not.
The GA gives symmetric spectra\cite{Fukushima05,Randeria05} if only
quasi-particle excitation is considered (incoherent excitations may
cause asymmetry\cite{Randeria05}). In contrast, the spectra calculated
by the VMC show asymmetry.\cite{CPChou06}

To discuss this point, here we calculate corrections to the results in
the former sections.
When these corrections are taken into account, the orthogonal relation,
Eq.~(\ref{eq:gammagamma}), may not be satisfied any more.
Therefore, in the following, we assume that the systems are almost
uniform; in the uniform limit the wave number is a good quantum number
due to the translational symmetry, and thus excited states are
orthogonal.
The next order corrections contain only site $i$ and $j$ similarly to
those in the hopping term.  We put general formulae in Appendix
\ref{sec:highDOS}, and here only show a special case of
$n_{i\uparrow}=n_{i\downarrow}=n_i/2$,
$n_{ij\uparrow}=n_{ji\uparrow}=n_{ij\downarrow}=n_{ji\downarrow}\equiv
n_{ij}$, $\Delta_{ij}=\Delta_{ji}=\Delta_{ij}^*=\Delta_{ji}^*$. Then,
with
\begin{gather}
 A_{ij}\equiv
\left\{
\begin{array}{ll}
\frac{ n_{ij}^2+\Delta_{ij}^2 }{(1-\frac{n_i}{2})(1-\frac{n_j}{2})}
\quad &(i\neq j)
\\
0 & (i = j)
~,
\end{array}
\right.
\\
 \alpha_i \equiv \frac{ \frac{n_{i}}{2} }{ 1-\frac{n_{i}}{2} }
~,
\end{gather}
Eqs.~(\ref{eq:cpcp1}-\ref{eq:cpcp4}) are rewritten as
\begin{align}
\label{eq:cPcPhigh1}
\frac{ \langle  c_{j\uparrow}^\dagger P c_{i\uparrow} P \rangle_0 }
{ \langle  P^2 \rangle_0 }
& \approx
\sqrt{ g_{ii}^{t0} } \,
\langle  c_{j\uparrow}^\dagger  c_{i\uparrow}  \rangle_0
\, ( 1 + \alpha_j A_{ij} )
~,
\\
\label{eq:cPcPhigh2}
\frac{ \langle  c_{j\uparrow} P c_{i\uparrow}^\dagger P \rangle_0 }
{ \langle  P^2 \rangle_0 }
& \approx
\sqrt{ g_{ii}^{t0} } \,
 \langle  c_{j\uparrow}  c_{i\uparrow}^\dagger \rangle_0
 \, ( 1 - A_{ij} )
~,
\end{align}
\begin{align}
\label{eq:cPcPhigh3}
\frac{ \langle  c_{j\downarrow}^\dagger P c_{i\uparrow}^\dagger P \rangle_0 }
{ \langle  P^2 \rangle_0 }
& \approx
\sqrt{ g_{ii}^{t0} } \,
\langle  c_{j\downarrow}^\dagger c_{i\uparrow}^\dagger \rangle_0
\, ( 1 - \alpha_j A_{ij} )
~,
\\
\frac{ \langle  c_{j\downarrow} P c_{i\uparrow} P \rangle_0 }
{ \langle  P^2 \rangle_0 }
& \approx
\sqrt{ g_{ii}^{t0} } \,
\langle  c_{j\downarrow} c_{i\uparrow} \rangle_0
 \, ( 1 + A_{ij} )
~.
\label{eq:cPcPhigh4}
\end{align}
Since $A_{ij}\ge 0$ and $\alpha_i \ge 0$, the corrections are
positive for the electron removal, and negative for the addition.
For more careful analysis, we also need to check the normalization.
Including the corrections, Eqs.~(\ref{eq:cPc1}-\ref{eq:cPc4}) are rewritten as
\begin{align}
\label{eq:cPPchigh1}
\frac{ \langle  c_{j\uparrow}^\dagger P^2 c_{i\uparrow} \rangle_0 }
{ \langle  P^2 \rangle_0 }
& \approx
\langle  c_{j\uparrow}^\dagger  c_{i\uparrow}  \rangle_0
\, ( 1 - \alpha_i \alpha_j A_{ij} )
~,
\\
\label{eq:cPPchigh2}
\frac{ \langle  c_{j\uparrow} P^2 c_{i\uparrow}^\dagger \rangle_0 }
{ \langle  P^2 \rangle_0 }
& \approx
 \langle  c_{j\uparrow}  c_{i\uparrow}^\dagger \rangle_0
 \, ( 1 - A_{ij} )
~,
\end{align}
\begin{align}
\frac{ \langle  c_{j\downarrow}^\dagger P^2 c_{i\uparrow}^\dagger \rangle_0 }
{ \langle  P^2 \rangle_0 }
& \approx
\langle  c_{j\downarrow}^\dagger c_{i\uparrow}^\dagger \rangle_0
\, ( 1 - \alpha_j A_{ij} )
~,
\\
\frac{ \langle  c_{j\downarrow} P^2 c_{i\uparrow} \rangle_0 }
{ \langle  P^2 \rangle_0 }
& \approx
\langle  c_{j\downarrow} c_{i\uparrow} \rangle_0
 \, ( 1 - \alpha_i A_{ij} )
~.
\label{eq:cPPchigh4}
\end{align}
These corrections to the normalization are all negative, and they do not
seem to cancel the asymmetry in Eqs.~(\ref{eq:cPcPhigh1}-\ref{eq:cPcPhigh4}).
Therefore, these results suggest that the higher-order GA exhibits
asymmetric spectra whose electron addition spectra are smaller than the
removal.

\begin{figure}[h]
\includegraphics[width=8cm]{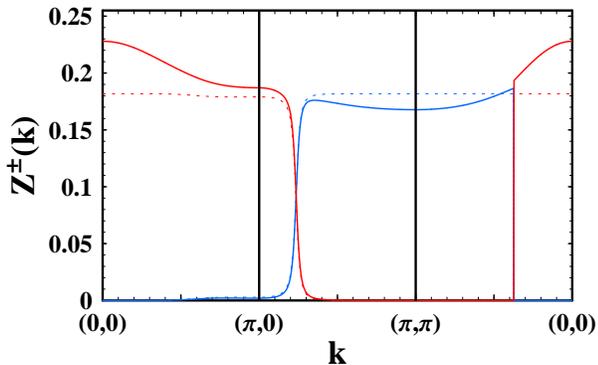}
\caption{
(Color online)
$Z^+(k)$ (blue lines) and $Z^-(k)$ (red lines) of a projected $d$-wave superconductor
by the conventional (dotted lines) and the generalized (solid lines) GA
with $t'=-0.3t$, $t''=0.2t$, $\Delta_{\rm v}=0.15t$, and
10\% hole concentration.
\label{fig:Zdwave}}
\end{figure}

This asymmetry is consistent with the VMC calculations for excitation
spectra by Chou {\it et al.}\cite{CPChou06}, and for spectral weights by
Bieri and Ivanov\cite{Bieri07} and Yang {\it et al.}\cite{HYYang07}
For more explicit comparison, we calculate the spectral weights,
\begin{eqnarray}
 Z^+({k}) & \equiv & | \langle k\sigma| c_{k\sigma}^\dagger |0\rangle |^2
\nonumber \\
&=&
\frac
{\langle P^2 \rangle_0}
{\langle \gamma_{k\sigma} P^2 \gamma_{k\sigma}^\dagger \rangle_0}
~
\left| \frac{\langle \gamma_{k\sigma} P c_{k\sigma}^\dagger P \rangle_0}
{\langle P^2 \rangle_0}\right|^2
,
\\
 Z^-({k}) & \equiv & | \langle k\sigma| c_{-k\bar\sigma} |0\rangle |^2
\nonumber \\
&=&
\frac
{\langle P^2 \rangle_0}
{\langle \gamma_{k\sigma} P^2 \gamma_{k\sigma}^\dagger \rangle_0}
~
\left| \frac{\langle \gamma_{k\sigma} P c_{-k\bar\sigma} P \rangle_0}
{\langle P^2 \rangle_0}\right|^2
,
\end{eqnarray}
and show them in Fig.~\ref{fig:Zdwave} for both the conventional and the
generalized GA.  Here, we include $t'$ and $t''$ in addition to
Eq.~(\ref{eq:dwaveBCS}) for a better correspondence to the high-$T_{\rm
c}$ superconductors.

In the case of the standard BCS theory, $Z^+=|u_k|^2$, $Z^-=|v_k|^2$.
Then, for each $k$-point below the Fermi level, one can find a
corresponding point $k'$ above the Fermi level such that $E_{k'}=E_{k}$,
$u_{k'}=v_k$, $v_{k'}=u_k$.  Then, summation of the contribution from
these two points to the spectra is unity for {\it both} addition and
removal spectra\cite{Tinkham72} because
$|u_k|^2+|u_{k'}|^2=|v_k|^2+|v_{k'}|^2=|u_k|^2+|v_k|^2 =1$.  Accordingly
the excitation spectra are symmetric.
The results of the conventional GA are $Z^+= g^t|u_k|^2$,
$Z^-=g^t|v_k|^2$; namely, the spectra are just renormalized by $g^t$, and
are symmetric as the standard BCS theory.
In contrast, by including the corrections to them, $Z^-$ decreases and 
$Z^+$ increases, which can cause the asymmetry in the spectra.
These $Z^\pm$ are consistent with the VMC results\cite{Bieri07,HYYang07}.
Note that $A_{ij}$ is finite even for $\Delta_{ij} = 0$, {\it i.e.,} the
Fermi sea also has the asymmetry, which is also consistent\cite{HYYang07}.
Similarly to the hopping term, 
Eqs.~(\ref{eq:cPcPhigh1}-\ref{eq:cPPchigh4})
are more accurate for small $|i-j|$.
Hence, the Fourier transformed results may include errors from
the summation over large $|i-j|$.
It will be checked in the future studies by including higher order terms.
Since this asymmetry appears as a deviation from the conventional GA,
it is rather small (especially near the Fermi level),
and does not look like what is seen in the STM experiment.

\subsection{Opposite asymmetry in projected $s$-wave superconductors}
\label{sec:swavedos}

We speculate that the origin of the asymmetry may not be so simple as
the intuition that {\it electron addition may be more difficult than
removal because electrons repel each other}.
Here, we show a counterexample against this simple scenario. That is to
say, projected $s$-wave superconductors can have the opposite asymmetry;
the electron addition spectra are larger than the removal.
Such projected $s$-wave superconductors may be realized if the pairing
interaction is isotropic because $d$-wave does not gain energy from
diagonal $J_{ij}$.
Even if $J_{ij}$ is finite only for nearest neighbors, the mean-field
approximation in very overdoped systems converges to extended $s$-wave
solutions.
To be more precise, this opposite asymmetry is related to finite on-site
pairing {\it before} the projection, $\langle c_{i\uparrow}^\dagger
c_{i\downarrow}^\dagger \rangle_0 \neq 0$ and not really related to the
symmetry of the gap.
Then, even for $d$-wave, inhomogeneity causes deviation from the
$d$-wave, and $\langle c_{i\uparrow}^\dagger c_{i\downarrow}^\dagger
\rangle_0$ can be nonzero in general.  Therefore, strongly disordered
$d$-wave superconductors could have similar properties.

To take $\Delta_{ii} \neq 0$ into account, we have to redo the
derivation from the beginning.
Then, $\Xi_i$ and $\lambda_i$ should be replaced by
\begin{gather}
 \Xi_i \approx \frac{(1-\frac{n_i}{2})^2 +|\Delta_{ii}|^2}{1 - n_i} ~,
\\
 \lambda_i \approx
\frac{n_i \big[ (1-\frac{n_i}{2})^2 + |\Delta_{ii}|^2 \big]}
{2(1-n_i)\big[(1-\frac{n_i}{2})\frac{n_i}{2}-|\Delta_{ii}|^2\big]} ~.
\end{gather}
In fact, this generalization makes analytical treatment very difficult,
and in the following we take only the leading order of the intersite
contractions.  Accordingly, its $\Delta_{ii}\rightarrow0$ limit
corresponds to the conventional GA (not the generalized GA in 
Sec.~\ref{sec:highaddrem}).

\begin{figure}[h]
\includegraphics[width=8cm]{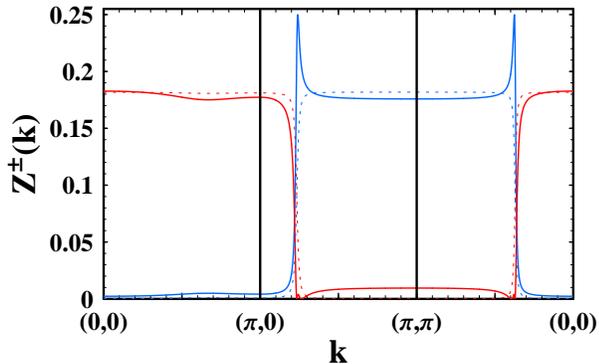}
\caption{(Color online)
$Z^+(k)$ (blue lines) and $Z^-(k)$ (red lines)
of a projected $s$-wave superconductor
with $t'=-0.3t$, $t''=0.2t$ $\Delta_{k}=0.15t$,
and 10\% hole concentration.
\label{fig:Zswave}}
\end{figure}

The most important matrix elements are
\begin{align}
\frac{ \langle  c_{i\downarrow}^\dagger P c_{i\uparrow}^\dagger P \rangle_0 }
{ \langle  P^2 \rangle_0 }
& \approx
 \frac{\sqrt{\lambda_i}}{\Xi_i} \,
\langle  c_{i\downarrow}^\dagger c_{i\uparrow}^\dagger \rangle_0
~,
\label{eq:ScPcP3}
\\
\langle  c_{i\downarrow} P c_{i\uparrow} P \rangle_0 &= 0
~.
\label{eq:ScPcP4}
\end{align}
Here, Eq.~(\ref{eq:ScPcP4}) is exact because of $P_{\rm G} c_{i\uparrow}
P_{\rm G}= c_{i\uparrow} P_{\rm G}$ and $P_{\rm G} c_{i\uparrow}^\dagger
c_{i\downarrow}^\dagger=0$.
The renormalization factor in Eq.~(\ref{eq:ScPcP3}) is obviously larger
than that in Eq.~(\ref{eq:ScPcP4}).  Hence, these matrix elements
suggest that the asymmetry is the opposite to that of $d$-wave.
We have also calculated other matrix elements, and after the Fourier
transform, $Z^\pm$ for a uniform system are obtained as plotted in
Fig.~\ref{fig:Zswave}.

In comparing Figs.~\ref{fig:Zdwave} and \ref{fig:Zswave}, we should keep
in mind that a better approximation is used for Fig.~\ref{fig:Zdwave},
and the asymmetry as in Fig.~\ref{fig:Zdwave} does not appear in
Fig.~\ref{fig:Zswave}.  Nevertheless, the characteristic asymmetry near
the Fermi surface in Fig.~\ref{fig:Zswave} is very strong, and may
remain even in calculation with a better precision.
Most likely, Eqs.~(\ref{eq:ScPcP3},\ref{eq:ScPcP4}) mainly contribute to
the asymmetry because the vicinity of the Fermi level changes most
dramatically.

\subsection{Physical consideration for the asymmetries}

For the projected $s$-wave superconductors, the
physical origin of the asymmetry may be understood as follows.  We have
been using terms ``addition'' and ``removal'' but these are in fact
named from the ground state's view.  If one takes the complex conjugate,
this addition (removal) matrix elements can be regarded as removal from
(addition to) an excited state.  Let us adopt the excited states' view
for a while.  In the $s$-wave BCS superconducting state
(before the projection), a Cooper pair may be formed more
or less on-site, which is a resonance of the doubly occupied state and
the empty state.
When $c_{i\sigma}$ is operated to this wave function, it chooses the the
``originally doubly occupied'' state.  Then, in
$Pc_{i\sigma}|\Psi\rangle_0$, the opposite spin state, $i\bar\sigma$, is
{\it occupied} with high probability.  Accordingly, it is and easy to
remove $i\bar\sigma$ electron.
In contrast, for $Pc_{i\sigma}^\dagger|\Psi\rangle_0$, it is impossible
to add $i\bar\sigma$ electron.
Finally, let us turn back to the ground state's view and review the
arguments above. Then, {\it the removal is difficult, but the addition
is easy}.

\begin{figure}[h]
\includegraphics[width=6cm]{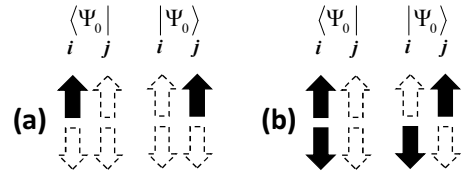} \caption{
 Configurations in $|\Psi_0\rangle$.
\label{fig:spinZ}}
\end{figure}

The asymmetry in the $d$-wave superconductors and the Fermi sea needs
more consideration because it appears by higher order correlations.
Fig.~\ref{fig:spinZ}(a) shows a configuration in the ket $|\Psi\rangle_0$
contributing to $\langle c_{j\uparrow} P c_{i\uparrow}^\dagger P
\rangle_0$.  The first term in Eq.~(\ref{eq:cPcPhigh2}) represents
direct correlation between $c_{j\uparrow}$ and
$c_{i\uparrow}^\dagger$.  The second term comes from the repulsive
correlation between down holes, which reduces weight of this
configuration.
On the other hand, for $\langle c_{i\uparrow}^\dagger P c_{j\uparrow}
P \rangle_0$, both configurations (a) and (b)
in $|\Psi\rangle_0$ contribute.  However, when electron density is high,
(b) is dominant because empty sites are rare.  Then,
correlation between down holes increases the weight of (b).  Since
this effect appears only at high density, the second term in
Eq.~(\ref{eq:cPcPhigh1}) accompanies the factor $\alpha_i$.

\section{Global constraint (quasi-canonical GA)}
\label{sec:global}

In the former sections, we have used the local constraint.  However, if one
needs the GA as an approximate method of the VMC, it may be preferred to
require the usual canonical constraint, {\it i.e.}, the total particle
number constraint for each of up and down spins,
\begin{equation}
 \sum_i \langle \hat{n}_{i\sigma}\rangle
= N_{\rm \sigma}^{\rm after}
~,
\label{eq:globalconstr}
\end{equation}
where $ N_{\rm \sigma}^{\rm after}$ is the total number of $\sigma$
electrons after the projection.  Although one takes $ N_{\rm
\sigma}^{\rm after} = \sum_i n_{i\sigma}$ in the usual canonical GA, the
particle numbers before and after the projection can be different in
general.  In fact, in the VMC, they are different; the particle number
projection $P_{N_{\rm \sigma}^{\rm after}}$ is usually applied together
with $P_{\rm G}$, and the chemical potential of $|\Psi_0 \rangle$ is
more like a variational parameter and does {\it not} control the
particle number {\it after} the projection.
In the following we use notation, 
\begin{equation}
  n_\sigma^{\rm after}\equiv N_\sigma^{\rm after}/N_{\rm L} ~,
\quad
 n^{\rm after} \equiv n_\uparrow^{\rm after}+n_\downarrow^{\rm after} ~,
\end{equation} 
with $N_{\rm L}$ the total number of sites.  
Our purpose here is formulating a grand canonical GA that gives results
of the canonical scheme, by imposing Eq.~(\ref{eq:globalconstr}).

If the total spin moment is nonzero, it must be reasonable to choose the
spin-$z$ axis parallel to the global moment so that $\sum_i \langle
S^x_i \rangle_0 = \sum_i \langle S^y_i \rangle_0 =0$.
In that case, local $xy$ components, $\langle S^x_i\rangle_0$, $\langle
S^y_i\rangle_0$, may be finite in general.  Then, similarly to the
local-constraint formulation in Sec.~\ref{sec:finiteSxy}, $xy$
components of the local spin moments are renormalized differently from
their $z$ components.
The canonical scheme condition for the total spin moment restricts the
spin-$z$ renormalization factor to the vicinity of unity, but the other
directions are free from it.  We expect that this spin-rotational
asymmetric renormalization is a property from the canonical condition
and should exist even in exact calculation.
Furthermore, if the total spin moment is zero and local moments point to
various directions, we have no idea how to choose the $z$-axis.
Here, to avoid such complexity, we assume $\langle S^x_i \rangle_0 = \langle
S^y_i \rangle_0 = 0$, and as in the former sections $
\langle c^\dagger_{i\sigma} c_{j\bar{\sigma}}
\rangle_0 = \langle c_{i\uparrow}^\dagger c_{i\downarrow}^\dagger
\rangle_0 =\langle c_{i\sigma}^\dagger c_{j\sigma}^\dagger \rangle_0 =
0$.

\subsection{Condition for fugacity factors}

To control the total particle numbers, we need a factor in the form
$\lambda_\sigma^{\frac12\sum_i \hat{n}_{i\sigma}}$, namely, the fugacity
factors $\lambda_{\sigma}$ do not have the site index.
Accordingly, the projected wave function is defined as $|\Psi\rangle = P
|\Psi_0 \rangle$ with $P \equiv \prod_i
\lambda_{\uparrow}^{\frac12\hat{n}_{i\uparrow} }
\lambda_{\downarrow}^{\frac12 \hat{n}_{i\downarrow}}
(1-\hat{n}_{i\uparrow} \hat{n}_{i\downarrow}) $.

The formula for $\langle \hat{n}_{i\sigma} \rangle$ has the same form as
Eq.~(\ref{eq:nrenormlocal}), and only $\lambda_\sigma$ is different,
{\it i.e.},
\begin{eqnarray}
 \langle \hat{n}_{i\sigma} \rangle
 \approx 
\frac
{
\lambda_{\sigma}
( 1 - {n}_{i\bar{\sigma}} )
}
{\Xi_i} \, {n}_{i\sigma}
~.
\label{eq:nrenorm}
\end{eqnarray}
Note that $\Xi_i$ is still site-dependent because it contains local
electron densities.
By inserting it into Eq.~(\ref{eq:globalconstr}), we obtain
\begin{equation}
\lambda_\sigma  \sum_i
\frac
{
 1 - {n}_{i\bar{\sigma}}
}
{\Xi_i} \, n_{i\sigma} 
 =  N_{\rm \sigma}^{\rm after}
~.
\label{eq:forlambda}
\end{equation}
In inhomogeneous systems, $\lambda_\sigma$ is solved numerically from
Eq.~(\ref{eq:forlambda}) in general.  An important point of this uniform
fugacity approach is that the local electron density is also
renormalized as in Eq.~(\ref{eq:nrenorm}), and $\langle\hat{n}_{i\sigma}
\rangle \neq \langle \hat{n}_{i\sigma} \rangle_0 $ in general.  When
$\lambda_\sigma$ is solved and inserted into Eq.~(\ref{eq:nrenorm}), the
corrections to $\langle \hat{n}_{i\sigma} \rangle$ are of the second
order of intersite contractions as will be explicitly shown in
Sec.~\ref{sec:beyond}.

The local spin-$z$ component is renormalized as
\begin{equation}
  \langle S^z_i \rangle
 \approx 
\frac12
\sum_{\sigma=\pm1}
\sigma
\frac
{
\lambda_{\sigma}
( 1 - {n}_{i\bar{\sigma}} )
}
{\Xi_i} \, {n}_{i\sigma}
~,
\label{mrenorm}
\end{equation}
where symbols $\uparrow,\downarrow$ and $+1,-1$ are interchangeably used.

\subsection{Hopping term}
The Gutzwiller renormalization factor of the hopping term is given by
\begin{equation}
\frac{\langle
c_{i\sigma}^\dagger c_{j\sigma}
\rangle}
{\langle
c_{i\sigma}^\dagger c_{j\sigma}
\rangle_0}
\approx
\frac{
 \lambda_{\sigma}
 (1-n_{i\bar{\sigma}}) (1-n_{j {\bar{\sigma}}}) }
{ \Xi_i \Xi_j }
\equiv
g_{ij\sigma}^{t0}
~.
\label{eq:gt}
\end{equation}
Next order corrections to this formula involves another site, which may
make important contribution for second or third neighbor hopping in some
systems.  Using $a_{i\sigma} \equiv (1-\lambda_{\bar\sigma})
(1-n_{i\sigma}) + \lambda_\sigma n_{i\sigma}$, it is written as
\begin{equation}
\langle  c_{i\uparrow}^\dagger c_{j\uparrow}  \rangle \approx
g_{ij\uparrow}^{t0}
\left(
n_{ij\uparrow}+\sum_l
\frac{ a_{l\downarrow} n_{il\uparrow} n_{lj\uparrow}
 -a_{l\uparrow} \Delta_{il}^*\Delta_{jl}    }{\Xi_l}
\right).
\end{equation}
The corrections to $\lambda_{i\sigma}$ and $\Xi_i$ affect only from third
order and not relevant to the equation above.
Note that $a_{i\sigma}$ goes to zero in the uniform limit with
$ n_\sigma^{\rm after} = n_{\sigma} $.

In the uniform systems, by omitting irrelevant site indices and
using $ R_\sigma \equiv { n_\sigma^{\rm after}
}/{ n_{\sigma} }$, we obtain
\begin{gather}
 \lambda_\sigma \approx
 \frac{ R_{\sigma}(1-n_\sigma) } { 1-n^{\rm after}  }
= \frac{ R_{\sigma}(1-n_\sigma) }
 { 1-R_{\uparrow}n_{\uparrow} -R_{\downarrow}n_{\downarrow}  }
~,
\label{eq:globalunilambda}
\\
g_{\sigma}^{t0}=
\frac{ R_\sigma (1-n^{\rm after}) }{1-n_\sigma}
=\frac{R_\sigma^2}{\lambda_\sigma}
~.
\label{eq:globalunigt}
\end{gather}

\subsection{Exchange term with zero total spin-$z$ component and
$\lambda_\uparrow=\lambda_\downarrow$}

The general formulae for the exchange interaction term are too lengthy
to present here.  Our main interest is in systems with zero total
spin-$z$ component and $\lambda_\uparrow=\lambda_\downarrow$ which
includes such as non-magnetic systems, antiferromagnets, and stripes.
Hence, for simplicity, we restrict ourselves to this case in the
following except for Sec.~\ref{sec:ferro} that treats ferromagnetic
systems.  The generalization to nonzero total spin-$z$ component is
straightforward but one has to work with more complexities.

When $\lambda_\uparrow=\lambda_\downarrow\equiv\lambda$,
Eq.~(\ref{mrenorm}) is reduced to
\begin{equation}
  \langle S^z_i \rangle
 \approx 
\frac{\lambda}{\Xi_i} \, m_i
~.
\end{equation}

For the exchange interaction term $\langle {\bf S}_{i} \cdot {\bf S}_{j}
\rangle $, we take up to the second order of intersite contractions.
Assuming that $m_i$ is small, namely, $m_i = O(n_{ij\sigma})= O(\Delta_{ij})$, we obtain
\begin{eqnarray}
&& \langle S^z_{i} S^z_{j}\rangle
\nonumber \\
& & \approx
\frac{g^{s}_{ij}}{4} \Big(
4 m_{i} m_{j}
-|n_{ij\uparrow}|^2-|n_{ij\downarrow}|^2-|\Delta_{ij}|^2-|\Delta_{ji}|^2
\Big)
\nonumber \\
 && =
g^{s}_{ij} \langle S^z_{i} S^z_{j}\rangle_0
~,
\label{eq:globalszsz0}
\end{eqnarray}
\begin{eqnarray}
 \langle
S^x_{i} S^x_{j}+S^y_{i} S^y_{j}
\rangle
& \approx &
- g^{s}_{ij}
{\rm Re}\big[n_{ij\uparrow}n_{ji\downarrow}
+\Delta_{ij}^* \Delta_{ji}
\big]
\nonumber \\
& =&
g^{s}_{ij}
\langle
S^x_{i} S^x_{j}+S^y_{i} S^y_{j}
\rangle_0
~,
\label{eq:glsxsyrenorm}
\end{eqnarray}
where $ g^{s}_{ij} \equiv {\lambda^2 }{(\Xi_i \Xi_j)^{-1}} $.  This is
the result of the conventional GA.  However, note that, if $m_{i}$ is of
the order of unity, terms such as $n_{ij\sigma}^2 m_{i}$, $\Delta_{ij}^2
m_{i}$, have about the same order of contribution as $n_{ij\sigma}^2$,
$\Delta_{ij}^2$, and formulae above should be modified as derived below.

\subsection{Beyond the ``conventional'' GA}
\label{sec:beyond}

When $m_i \sim n_{i\sigma}$, terms neglected in the previous derivation
may grow, and we need to redo the derivation from the beginning.  It is
known that expectation values by Gutzwiller-projected states can be
written in the form of a linked cluster expansion.\cite{Gutzwiller63}
The terms we need here includes contribution from clusters one-site
larger than those in the previous derivation.

We relegate detailed derivation to Appendix \ref{sec:hiorder}, and only
show final results here.  The renormalization of the particle densities
is given by
\begin{widetext}
\begin{eqnarray}
 \langle \hat{n}_{i\uparrow} \rangle & \approx &
  \frac{ \lambda (1-n_{i\downarrow}) }{\Xi_i} n_{i\uparrow}
\left (1 - \sum_{l\neq i}
\frac{\Xi^{(2)}_{il}}{\Xi_{i} \Xi_{l}}
\right) 
\nonumber\\&&
+\sum_{l\neq i} \frac{\lambda}{\Xi_i\Xi_l}  \bigg\{ 
 a_{l\downarrow}
\Big[ ( 1 - n_{i\downarrow}) |n_{il\uparrow}|^2 + n_{i \uparrow} |\Delta_{li}|^2 \Big]
- a_{l \uparrow}
\Big[ n_{i\uparrow} |n_{il\downarrow}|^2 + ( 1 - n_{i\downarrow} ) |\Delta_{il}|^2 \Big]
\bigg\},
\label{eq:globalnrenorm}
\end{eqnarray}
\begin{equation}
\Xi^{(2)}_{lm} \equiv
- a_{l\uparrow} a_{m \uparrow} |n_{lm \downarrow}|^2
- a_{l \downarrow} a_{m \downarrow} |n_{lm \uparrow}|^2
+ a_{l \downarrow} a_{m \uparrow} |\Delta_{lm}|^2
+ a_{l\uparrow} a_{m \downarrow} |\Delta_{ml}|^2
~,
\end{equation} 
where $ a_{l\sigma} = (1-\lambda)(1-n_{l\sigma}) + \lambda n_{l \sigma}
$.  The formula for $\langle \hat{n}_{i\sigma} \rangle$ is obtained by
replacing as $\uparrow \Leftrightarrow \downarrow$ and $\Delta_{il}
\Rightarrow - \Delta_{li}$.  Then, the new equation to determine
$\lambda$ is given by $\sum_i \langle \hat{n}_{i\sigma}\rangle= N^{\rm
after}$.  The solution can be written as
$\lambda\approx\lambda^{(0)}+\lambda^{(2)}$, where $\lambda^{(0)}$ is
$\lambda$ determined by Eq.~(\ref{eq:forlambda}), and $\lambda^{(2)}$ is
the correction to it represented by
\begin{eqnarray}
\lambda^{(2)} &=& \bigg\{
\sum_i \Xi_i^{-2}
 (1-n_{i\uparrow})(1-n_{i\downarrow})
\Big[ n_{i\uparrow}(1-n_{i\downarrow})
+n_{i\downarrow}(1-n_{i\uparrow}) \Big]
\bigg\}^{-1}
 \nonumber\\ &&
\times
\sum_i \sum_{l\neq i} \frac{\lambda^{(0)}}{\Xi_i\Xi_l} 
\Bigg\{ \frac{ n_{i\uparrow}(1-n_{i\downarrow})
             +n_{i\downarrow}(1-n_{i\uparrow})}{\Xi_i}
~ \Xi^{(2)}_{il}
- a_{l\downarrow}
\Big[ (1-2n_{i\downarrow}) |n_{il\uparrow}|^2   - (1-2n_{i \uparrow}) |\Delta_{li}|^2 \Big]
\nonumber\\&& \qquad\qquad\qquad\qquad\qquad\qquad\qquad\qquad\qquad\qquad\qquad
- a_{l \uparrow}
\Big[ (1-2n_{i\uparrow})   |n_{il\downarrow}|^2 - (1-2n_{i\downarrow})|\Delta_{il}|^2 \Big]
\Bigg\}.
\label{eq:lambda2}
\end{eqnarray}
Here, every $\lambda$ is replaced by $\lambda^{(0)}$
in the r.h.s.
Using this new $\lambda$, we can calculate spin terms,
\begin{eqnarray}
\langle S^z_{i} \rangle 
& \approx &
\frac{\lambda m_i}{\Xi_i}
\left(1 - \sum_{l\neq i} \frac{\Xi^{(2)}_{{i}{l}}}{\Xi_{i} \Xi_{l}}\right)
+\sum_{l\neq i}m^{(2)}_{il},
\quad m^{(2)}_{il} \equiv
  \frac{\lambda}{2\Xi_i\Xi_l} \Big[
   a_{l\downarrow} \left(\, |n_{il\uparrow}|^2 + |\Delta_{li}|^2 \,\right)
 -  a_{l\uparrow} \left(\, |n_{il\downarrow}|^2+ |\Delta_{il}|^2 \,\right)
\Big],
\end{eqnarray}
%
%
\begin{eqnarray}
\langle S^z_{i} S^z_{j} \rangle - \langle S^z_{i}\rangle \langle S^z_{j} \rangle
 &\approx&
\frac{\lambda^2 m_i m_j}{\Xi_i\Xi_j}
\left[1 + \frac{\Xi^{(2)}_{{i}{j}}}{\Xi_{i} \Xi_{j}}\right]
+ \frac{\lambda^2}{\Xi_i\Xi_j}
\Big(\langle S^z_{i} S^z_{j} \rangle_0 - m_i m_j\Big)
 -  \frac{\lambda m_j}{\Xi_j} m^{(2)}_{ij}
 -  \frac{\lambda m_i}{\Xi_i} m^{(2)}_{ji}
~,
\end{eqnarray}
Here, $\lambda$ in the second order terms can be replaced by
$\lambda^{(0)}$.  Note that the contribution that involves a third site
$l$ in $\langle S^z_{i}\rangle$ cancels that in $\langle S^z_{i}
S^z_{j}\rangle$ by the subtraction of $\langle S^z_{i}\rangle \langle
S^z_{j} \rangle$.  There is no correction of this order for $ \langle
S^x_{i} S^x_{j}+S^y_{i} S^y_{j} \rangle$ in Eq.~(\ref{eq:glsxsyrenorm}).

Although several authors\cite{TCHsu90,Sigrist94,Ogata03} formulated
improved canonical GAs by taking nearest-neighbor correlations similarly
to ours, our result is different from any of them even if we neglect the
second- and the third-neighbor terms.  The origin of this discrepancy is
not clear at present.

\subsection{Antiferromagnets}
As an explicit example, we show the formulae for the square lattice
antiferromagnet.  For periodic systems, we can restrict the summation
over the site index $i$ to only inside of the unit cell.  In the
presence of the antiferromagnetic moments, $n_{ij}$ between second- or
third-neighbor pairs may be comparable to or larger than that of the
nearest neighbor pairs.  To take into account these terms, we define
$n_{ij\sigma}=\chi, \chi', \chi''$, $\Delta_{ij}=\Delta_{ji}= \Delta,
\Delta', \Delta''$, for the nearest, the second, the third neighbor
pairs, and assume these are real numbers.  In addition,
$n_{i\uparrow}=n_A$ and $n_{i\downarrow}=n_B$ for A-sublattice,
$n_{i\uparrow}=n_A$ and $n_{i\downarrow}=n_A$ for B-sublattice, and then
$m=(n_A-n_B)/2$.  By omitting irrelevant site indices,
\begin{equation}
\lambda^{(0)} = \frac{n^{\rm after}(1-n_A)(1-n_B)}{(1-n^{\rm after})(n-2 n_A n_B)}
~,\qquad
 a_A=(1-\lambda)(1-n_A) + \lambda n_A
~,\qquad
 a_B=(1-\lambda)(1-n_B) + \lambda n_B
~,
\end{equation}
\begin{gather}
g^{t0} = \frac{ (1-n^{\rm after})n^{\rm after} }{n-2n_A n_B}
~,\quad
\Xi = (1-n_A)(1-n_B) + \lambda (n - 2 n_A n_B)
~,\quad
 \Xi^{(2)}_{\rm n.n.} = -2 \, a_A \, a_B \, \chi^2 + (a_A^2 + a_B^2) \Delta^2
~,
\\
 \Xi^{(2)}_{\rm 2nd} =
 - (a_A^2 + a_B^2) (\chi')^2  + 2 \, a_A \, a_B \, (\Delta')^2
~, \qquad
 \Xi^{(2)}_{\rm 3rd} =
 - (a_A^2 + a_B^2) (\chi'')^2 + 2 \, a_A \, a_B \, (\Delta'')^2
~,
\end{gather}
\begin{align}
 \lambda^{(2)} = & 
\frac{4~\lambda}{(1-n_A)(1-n_B)(n-2 n_A n_B)}
\Bigg\{
\frac{(n-2 n_A n_B)}{\Xi}
\left( \Xi^{(2)}_{\rm n.n.}
   +   \Xi^{(2)}_{\rm 2nd}
   +   \Xi^{(2)}_{\rm 3rd} \right) \nonumber\\ & \qquad\qquad\qquad\qquad
+ \big[a_A (1-2n_B)+ a_B (1-2n_A)\big]
  \big[-\chi^2 +(\Delta')^2 +(\Delta'')^2 \big]
 \nonumber\\ & \qquad\qquad\qquad\qquad
+ \big[a_A (1-2n_A)+ a_B (1-2n_B)\big]\big[\Delta^2 -(\chi')^2 -(\chi'')^2\big]
\quad \left. \Bigg\} \right|_{\lambda\rightarrow\lambda^{(0)}}
~,
\end{align}
\begin{equation}
\langle S^z_{A} \rangle = -\langle S^z_{B} \rangle 
 \approx 
\frac{\lambda m}{\Xi}
\left(1 
- \frac{4\,\Xi^{(2)}_{\rm n.n.}}{\Xi^2 }
- \frac{4\,\Xi^{(2)}_{\rm 2nd}}{\Xi^2 }
- \frac{4\,\Xi^{(2)}_{\rm 3rd}}{\Xi^2 }
\right)
+ \frac{4\lambda(2\lambda-1)m}{\Xi^2}
\Big[ \chi^2 + \Delta^2 -(\chi')^2 - (\Delta')^2 -(\chi'')^2 - (\Delta'')^2
\Big]
~,
\end{equation}
\begin{equation}
\langle S^z_{A} S^z_{B} \rangle - \langle S^z_{A}\rangle \langle S^z_{B} \rangle
 \approx
\frac{- \lambda^2 m^2}{\Xi^2}
\left[1 + \frac{\Xi^{(2)}_{\rm n.n.}}{\Xi^2}\right]
-\frac{\lambda^2}{2\Xi^2}( \chi^2+\Delta^2 )\left[ 1
-  \frac{4 m^2}{\Xi}(2\lambda-1)
\right]
~,
\label{globalAFss}
\end{equation}
where Eq.(\ref{globalAFss}) is for a nearest neighbor pair.
In general, $\Delta_{ij}\neq \Delta_{ji}$ may
occur; in that case these equations need modification with a little more
complexities.
\end{widetext}

\subsection{Ferromagnets}
\label{sec:ferro}

Here, we show show a remarkable difference from the local constraint.
That is, ferromagnets are renormalized very differently from
antiferromagnets in contrast to results by the local constraint in
Sec.~\ref{sec:localex}.
For ferromagnetic wave functions without superconductivity, we can set
$\Delta_{ij}=0$, and $N_\sigma^{\rm after}= \sum_i \langle
\hat{n}_{i\sigma} \rangle_0$.
Then, in the uniform cases, this GA with the global constraint
is equivalent to the one with the local-constraint.
Therefore, by setting $m_i=m$ in formulae in Sec.~\ref{sec:local},
we obtain those for the ferromagnets,
\begin{equation}
\langle S^z_{i} S^z_{j}\rangle
 \approx
 m^2   -\frac14
 \left[
 |n_{ij\uparrow}|^2 \frac{(1-2m)^2}{(1-n_{\uparrow})^2}
+|n_{ij\downarrow}|^2 \frac{(1+2m)^2}{ (1-n_{\downarrow})^2}
\right],
\label{eq:ferroszsz}
\end{equation}
\begin{gather}
 \langle  S^x_{i} S^x_{j}+S^y_{i} S^y_{j} \rangle
 =
g^s \langle S^x_{i} S^x_{j}+S^y_{i} S^y_{j} \rangle_0 ~,
\label{eq:ferrosxsx}
\\
g^s \equiv \frac1{ (1-n_{\uparrow})(1-n_{\downarrow}) } ~,
\quad
g^t_{\sigma} = \frac{1-n}{1-n_\sigma}
~.
\label{eq:ferrogs}
\end{gather}
Many of the formulae derived with the global constraint in this
section contain $a_\sigma$, but it goes to zero in this
ferromagnetic limit.  It is consistent with no appearance of
$a_\sigma$ in the local-constraint formulation.

In fact, the renormalization for the spin-$z$ component represented by
Eq.~(\ref{eq:ferroszsz}) is different from the one derived by Zhang {\it
et al.}\cite{FCZhang88}\ using a probability argument of the canonical GA,
namely,
$ \langle S^z_{i} S^z_{j}\rangle=
g^s \langle S^z_{i} S^z_{j}\rangle_0
$ with $g^s$ defined by Eq.~(\ref{eq:ferrogs}).
However, we speculate that our result is more
reasonable because spin moment term $m^2$ is not renormalized; the
canonical constraint prevents the spin-$z$ component from growing
larger, in contrast to the antiferromagnetic moments, which are not
bound by the canonical constraint.
It may be clearer if we take the limit of small $m$ for
Eq.~(\ref{eq:ferroszsz}),
\begin{eqnarray}
\langle S^z_{i} S^z_{j}\rangle &\approx & 
m^2 -\frac{g^{s}_{ij}}{4} \Big(
|n_{ij\uparrow}|^2+|n_{ij\downarrow}|^2
\Big)
\nonumber \\
 & = &
 m^2 +
g^{s}_{ij} \Big(\langle S^z_{i} S^z_{j}\rangle_0
- m^2 \Big)
~.
\end{eqnarray}
and compare with Eq.~(\ref{eq:globalszsz0}) for antiferromagnets.

\subsection{Effect of $N^{\rm after}\neq N^{\rm before}$}

Projections reduce the Hilbert space.  Hence, many wave functions may be
equivalent to each other after the projection even if they are different
before the projection.  Here, we demonstrate it explicitly by the particle
number projection.  Let us start from uniform non-magnetic cases.
Define two BCS states,
\begin{gather}
|\Psi_0 \rangle\equiv
\prod_k (u_k + v_k c_{k\uparrow}^\dagger c_{-k\downarrow}^\dagger)
 |0\rangle
~,
\\
|\Psi'_0\rangle \equiv \tilde{\lambda}^{\hat{N}/2}|\Psi_0 \rangle
= \prod_k
(u_k + v_k \tilde{\lambda}
c_{k\uparrow}^\dagger c_{-k\downarrow}^\dagger) |0\rangle
~,
\label{eq:fugBCS}
\end{gather}
where $\hat{N} \equiv \sum_{i\sigma} \hat{n}_{i\sigma}$.
Under the particle number projection $P_N$,
\begin{equation}
  P_N |\Psi'_0 \rangle
= \tilde{\lambda}^{N/2} P_N |\Psi_0 \rangle
\propto P_N |\Psi_0 \rangle
~.
\end{equation}
Namely, wave functions $ P_N |\Psi_0 \rangle$ and
$P_N |\Psi'_0 \rangle$  are equivalent whereas
$|\Psi_0 \rangle$ and $|\Psi'_0 \rangle$ are nonequivalent.
At a sight, the quasi-particle excited states of these two BCS states
look different because $\gamma_{k\sigma}^\dagger$
does not commute with $\tilde{\lambda}^{\hat{N}/2}$.
However, $c_{k\sigma}^\dagger$ and $c_{k\,-\sigma}$ in
$\gamma_{k\sigma}^\dagger$ in fact yield the same state, and thus
$P_N \gamma_{k\sigma}^\dagger |\Psi_0 \rangle$ and
$P_N \gamma_{k\sigma}^{\prime\dagger} |\Psi'_0\rangle$
are equivalent, where $\gamma_{k\sigma}^{\prime\dagger}$ is a quasi-particle
operator for $|\Psi'_0\rangle$.

Therefore, even if the average particle number of $|\Psi_0 \rangle$ is
not $N^{\rm after}$, one can make that of $|\Psi'_0 \rangle$ equal to 
$N^{\rm after}$ by choosing $\tilde{\lambda}$ to satisfy
\begin{equation}
 N^{\rm after}=\sum_k
\frac{2\tilde{\lambda}^2 |v_k|^2}{|u_k|^2 + \tilde{\lambda}^2 |v_k|^2}
~.
\end{equation}
then, using $|\Psi'_0 \rangle$, the GA can be applied with the local constraint
$
\langle \hat{n}_{i\sigma} \rangle =
\langle \hat{n}_{i\sigma} \rangle_0
$.  Accordingly, we can use convenient properties derived in
Secs.~\ref{sec:local} and \ref{sec:localfug-xstates}.

Such a transformation to relate the global constraint to the local one
may be possible also for inhomogeneous systems, but there seems to be a
problem.  The particle numbers can be controlled by fugacity factors
$\prod_{i\sigma}\tilde{\lambda}_{i\sigma}^{\hat{n}_{i\sigma}/2}$ as
Eq.~(\ref{eq:fugBCS}).  One can choose $\tilde{\lambda}_{i\sigma}$ in
$|\Psi'_0 \rangle$ to satisfy $ \langle \hat{n}_{i\sigma} \rangle =
\langle \hat{n}_{i\sigma} \rangle_0 $.  Then,
$c_{i\sigma}^\dagger$ is replaced by
$\tilde{\lambda}_{i\sigma}^{\frac12} c_{i\sigma}^\dagger$ as well as
$c_{i\sigma}$ is replaced by $\tilde{\lambda}_{i\sigma}^{-\frac12}
c_{i\sigma}$ (in annihilating the electron, the fugacity factor caused
by the creation should be canceled).
Accordingly, $|\Psi'_0 \rangle$ is a rather strange wave function
because the quasi-particles may not satisfy the fermion commutation
relation.  Then one may need to redefine a proper quasi-particle set for
$|\Psi'_0 \rangle$.
Furthermore, since the quasi-particle operators do not commute with
the fugacity's operator, definition of the excited states depends on
their order, in contrast to the uniform cases.

We have originally speculated that the difference between the particle
numbers before and after the projection may cause such asymmetry of the
spectra as discussed in the latter part of
Sec.~\ref{sec:localfug-xstates}.
Let us look again at
Eqs.~(\ref{eq:localciPciP1},\ref{eq:localciPciP2}).
Note that the electron removal matrix element is proportional to
$n_{i\sigma}$ (density of the $i\sigma$ electrons), while the addition is
to $1-n_{i}$ (density of the {\it empty} sites).
Nevertheless, it is compensated by the fugacity factor and the
renormalization factors for the removal and the addition are the same.
Then, one may speculate that some asymmetry may appear if we destroy
this balance by changing the fugacity factors.
However, according to our analysis here,
the particle number difference does not have much effect,
and it does not cause any asymmetry at least in the uniform systems.

\section{Spin-independent constraint}
\label{sec:cano-gcano}

At present, our main interest is in the GAs with spin-dependent
constraints in the former sections because they seem to be more
convenient to investigate antiferromagnets, stripe state, impurity
systems, and so on.
However, in systems with a more complicated spin configuration, the GAs
with spin-{\it independent} constraints may be useful.  Therefore, here
we work on it, but only take the leading order with respect to the
intersite contractions.

\subsection{Local constraint}
\label{sec:cano-gcanolocal}

A grand canonical GA with a spin-independent constraint,
\begin{equation}
 \langle \hat{n}_{i\uparrow}+\hat{n}_{i\downarrow} \rangle = n_i ~,
\end{equation}
was introduced by Wang {\it et al.}\cite{QHWang06}
In non-magnetic cases ($n_{i\uparrow}=n_{i\downarrow}$), this is identical
to the GA with the spin-dependent constraint in Secs.~\ref{sec:local}
and \ref{sec:localfug-xstates}.
However, in magnetic cases, the results of these two GAs are different.
Accordingly, the ferromagnetic homogeneous limit\cite{WHKo07} with
the spin-independent constraint is not equivalent to that of the
canonical GA, but is reduced to the GA for charge--canonical
spin--grand-canonical functions explained in the next subsection.

Since the formulae for $g^t$ and $g^s$ have been already derived in
Refs.~\onlinecite{QHWang06,WHKo07}, here we derive them in a slightly
more general form by assuming that $\langle S^x_i \rangle$ and $\langle
S^y_i\rangle$ are finite.  By replacing $\lambda_{i\sigma}$
by $\lambda_{i}$ in the derivation in Sec.~\ref{sec:finiteSxy} and using
${\cal S}^{\pm}_{i} \equiv \langle S^{\pm}_{i} \rangle_0$, we obtain
\begin{gather}
 \Xi_i  \approx
\frac{\xi_i}{1-n_i}
~, \quad
\xi_i \equiv (1-n_{i\uparrow}) (1-n_{i\downarrow}) - |{\cal S}^+_i|^2
, \\
\lambda_i \approx
\frac
{ n_{i} \xi_i }
{ (1-n_i)\big[ n_i - 2 n_{i\uparrow}n_{i\downarrow}
   + 2 |{\cal S}^+_i|^2 \big] }
~,
\end{gather}
\begin{gather}
\frac{ \langle  c_{i\sigma}^\dagger c_{j\sigma}  \rangle }
    { \langle  c_{i\sigma}^\dagger c_{j\sigma}  \rangle_0 }
 \approx
g_{ij\sigma}^{t} = \sqrt{g_{ii\sigma}^{t} g_{jj\sigma}^{t}}
~,
\\
{g_{ii\sigma}^{t}} \equiv
\frac{ n_i(1-n_i)(1-n_{i\bar\sigma})^2}
{
\xi_i
[ n_i - 2 n_{i\uparrow}n_{i\downarrow} +2|{\cal S}^+_i|^2 ]
}
~,
\end{gather}
\begin{gather}
\langle {\bf S}_i \rangle
 \approx
\frac{\lambda_i}{\Xi_i} \, \langle {\bf S}_i \rangle_0
 = \sqrt{g^s_{ii}} \, \langle {\bf S}_i \rangle_0
~,
\\
\sqrt{g^s_{ii}} \equiv
\frac{ n_{i} }
{ n_i - 2 n_{i\uparrow}n_{i\downarrow} + 2|{\cal S}^+_i|^2 }
~,
\end{gather}
\begin{gather}
\langle {\bf S}_i \cdot {\bf S}_j \rangle
 \approx
\frac{\lambda_i}{\Xi_i} \frac{\lambda_j}{\Xi_j}
\, \langle {\bf S}_i \cdot {\bf S}_j \rangle_0
 = g^s_{ij} \,  \langle {\bf S}_i \cdot {\bf S}_j \rangle_0
~,
\\
g^s_{ij} \equiv \sqrt{ g^s_{ii} g^s_{jj} }
~.
\end{gather}
In this case, the derivation of $g^s$ is rather simple because ${\bf
S}_i$ is nonzero only when it is operated to states where site $i$ is
singly occupied.\cite{QHWang06} Note that $\langle {\bf S}_i \rangle
\parallel \langle {\bf S}_i \rangle_0$ is automatically satisfied, and
there is no complexity appeared in Sec.~\ref{sec:finiteSxy}.
However, in this formulation, projected quasi-particle excited states
are not orthogonal for magnetic systems in general.

\subsection{Global constraint
(charge--canonical spin--grand-canonical GA)}
\label{chargeqcano-spingcano}

It is possible that a wave function before the projection is an
eigenstate of the total particle number, but {\it not} any eigenstate of
the total spin. One can formulate a GA also for such systems.
In this case, the condition to impose is the canonical condition for
the total particle number,
$
\sum_{i\sigma} \langle \hat{n}_{i\sigma} \rangle = N^{\rm after}
~,
$
 namely,
\begin{equation}
 \lambda \sum_{i\sigma}
\frac
{
{n}_{i\sigma} ( 1 - {n}_{i\bar{\sigma}} )
}
{\Xi_i}=
 N^{\rm after}
~.
\label{eq:forlambda2}
\end{equation}
Here, $\lambda$ does not depend on $\sigma$.
The renormalization formulae for ${\cal S}^{\pm}_{i} = 0$ are the same
as those in Sec.~\ref{sec:global} if $\lambda_\sigma$ is replaced by
$\lambda$.  The generalization to ${\cal S}^{\pm}_{i} \neq 0$ is
straightforward.

The results for antiferromagnets are equivalent to those in
Sec.~\ref{sec:global}.  The limit to uniform ferromagnets without
superconductivity can be taken by setting $\Delta_{ij}=0$,
$N_\sigma^{\rm after}= \sum_i \langle \hat{n}_{i\sigma} \rangle_0$ and
dropping site indices,\cite{WHKo07}
\begin{equation}
g^t_\sigma =
 \frac{ (1-n_{\bar\sigma})(1-n)n}{(1-n_\sigma)(n-2n_\uparrow n_\downarrow)} ~,
\quad
g^s_\sigma =
 \left(\frac{ n }{n-2n_\uparrow n_\downarrow}\right)^2 ~.
\label{eq:spingrand}
\end{equation}
These are different from our quasi-canonical derivation in
Eqs.~(\ref{eq:ferroszsz},\ref{eq:ferrosxsx}).
In fact, in this spin--grand-canonical formulation, $g^s$ for
ferromagnets is the same as that for antiferromagnets.
This discrepancy is explained as follows: If the wave function before
the projection is an eigenstate of the total spin-$z$, then
the renormalization is represented by
Eqs.~(\ref{eq:ferroszsz},\ref{eq:ferrosxsx}).  If not, by
Eq.~(\ref{eq:spingrand}).
The Gutzwiller projection tends to magnify spin moments as explained in
Appendix \ref{sec:diffscheme}.  However, in the canonical scheme, only
$xy$ components of the spins are allowed to be enhanced because of the
canonical constraint.  On the other hand, the spin--grand-canonical case
is free from this constraint, and spins are renormalized more
isotropically.

\section{Summary and discussion}
\label{sec:summary}

We have derived various formulae using the grand canonical Gutzwiller
approximation with several different constraints imposed by the
fugacity factors for inhomogeneous magnetic systems.  The formulation
with the local particle number conservation yields more simple formulae.
On the other hand, the global particle number constraint is more
convenient in comparing with the VMC.

In Secs.~\ref{sec:localfug-xstates}, we have
discussed the asymmetry of the density of states.
Although the incoherent spectra are not taken into account in this
paper, we speculate that they appear at much higher energy scale than
the coherence peaks.
The conventional BCS theory tells us that the quasi-particle excitation
spectra are symmetric between positive and negative bias.  In contrast,
with the Gutzwiller projection, some asymmetry appears.
One may think that electron addition is always more difficult than
electron removal if repulsion between electrons is strong.  However, we
doubt if such simple intuition works, and speculate that the asymmetry
depends on the Hamiltonian.  As a counterexample, we have shown that the
projected $s$-wave superconductor may have the opposite
asymmetry. Namely, even with the strong repulsion, the addition spectra
can be larger than the removal.
We could be able to consider in this way.  Let us take two (normalized)
Gutzwiller-projected wave functions, $|\psi\rangle$ and $|\phi\rangle$.
Suppose they are the ground states of Hamiltonians, $H_\psi$ and
$H_\phi$, respectively.  
Furthermore, we assume that $|\psi\rangle$ and $|\phi\rangle$ are also excited
states of the other Hamiltonian, $H_\phi$ and $H_\psi$, respectively.
Then, $ \langle \phi| c_{i\sigma}^\dagger |\psi\rangle $
is an electron addition matrix element to the ground state of $H_\psi$.
However, if one takes its complex conjugate, it is
an electron removal matrix element to the ground state of $H_\phi$.
Note that $| \langle \psi| c_{i\sigma} |\phi\rangle |
=| \langle \phi| c_{i\sigma}^\dagger |\psi\rangle |$.
That is, if an electron addition matrix element for a Hamiltonian is
small, an electron {\it removal} matrix element for a different Hamiltonian is
also small.
Therefore, the asymmetry is most likely determined not only by the
projection, but also by the Hamiltonian.

Our ultimate purpose is to find good variational wave functions for
systems with strong on-site Coulomb repulsion.
Once the fugacity factors are introduced,
one projected wave function can be related to
a number of different {\it unprojected} wave functions,
each of which is accompanied with fugacity factors of each definition.
Therefore, one should probably choose a definition of fugacity factors
that matches their purpose.
We speculate that the projected optimized solution
is similar in any choice of the fugacity factors
if the calculation is done accurately enough.

There may be slight disagreement with the results by Ko {\it et al.}
That is, their comparison between the VMC and the GA seems to say that
the GA with the position- and spin-dependent constraint has larger errors
than that with the global constraint.  However, 
according to our estimation, the formulation with
the position- and spin-dependent constraint has much smaller errors.

To improve the approximation, one can use techniques of the series
expansion method, such as the finite cluster method for calculating
higher order terms.
One can also use the Pad{\'e} approximation for extrapolation if necessary,
but maybe it is enough just to neglect small terms without extrapolation.
Since this linked-cluster expansion can be done analytically, there is a
possibility to minimize the energy analytically in contrast to the
VMC.

\acknowledgments
The author thanks T.K.\ Lee for stimulating discussions.
Former collaboration with C.\ Gros, V.N.\ Muthukumar and B.\ Edegger was
also helpful in formulating this theory.
This work was supported by the National Science Council in Taiwan with
Grant no.95-2112-M-001-061-MY3. 
Part of the numerical calculations were performed in the FormosaII
Cluster in the National Center for High-performance Computing in Taiwan.

\appendix

\section{Difference between canonical and grand canonical scheme}
\label{sec:diffscheme}

Suppose $|\Psi_0^N \rangle$ is an eigenstate of the total particle number
$\hat{N} \equiv \sum_{i\sigma} \hat{n}_{i\sigma} $ with
$\hat{N} |\Psi_0^N \rangle = N |\Psi_0^N \rangle $.
Then, since $[P_{\rm G},\hat{N}]=0$,
\begin{equation}
 \hat{N} P_{\rm G}|\Psi_0^{N}\rangle =
 N P_{\rm G}|\Psi_0^{N}\rangle
~.\\
\end{equation}
Namely, in the canonical scheme, $P_{\rm G}$ does not change the
particle number.  On the other hand in the grand canonical scheme, a
wave function $|\Psi_0 \rangle$ is {\it not} an eigenstate of $\hat{N}$,
and the particle number is distributed with the distribution function,
\begin{equation}
 \rho^{(0)}_N \equiv \frac{\langle \Psi_0| P_N
|\Psi_0\rangle}{\langle \Psi_0|\Psi_0\rangle}\nonumber
~,
\end{equation}
where $P_N$ is an operator which projects onto terms with particle
number $N$.  Suppose $\rho^{(0)}_N$ is sharply peaked at a mean value
$N$, and fluctuation around it is of $O(\sqrt{1/N})$.  Then, $|\Psi_0
\rangle$ can be regarded as a wave function almost identical with
$|\Psi_0^N \rangle$ in the thermodynamic limit.  In contrast in the
projected case, the average particle number of $P_{\rm G}|\Psi_0\rangle$ is
in fact different from that of $P_{\rm G}|\Psi_0^N\rangle$.
When $N$ is large, an electron has more chance to meet another electron
on a certain site.  In other words, $P_{\rm G}$ excludes more states with
large $N$, and thus the peak position of the $N$ distribution is shifted to a
smaller value.
Such distribution change was explicitly estimated by
Edegger {\it et al.}\cite{Edegger05}\ as summarized below.
The distribution function after the projection,
\begin{equation}
{\rho}_N \equiv \frac{\langle \Psi_0|\,P_{\rm G} P_N P_{\rm G}
\,|\Psi_0\rangle}{\langle \Psi_0|P_{\rm G}P_{\rm G}|\Psi_0\rangle}~,
\end{equation}
can be related to that before by $\rho_N
= g_N \rho^0_N$ with
\begin{equation}
g_N \equiv C \frac{\langle \Psi_0|
P_{\rm G} P_N P_{\rm G} |\Psi_0\rangle}{\langle \Psi_0 |\,P_N\,|\Psi_0\rangle} \nonumber
~,
\end{equation}
where $C$ is a constant independent of $N$ coming from the normalization
of the wave functions.  The
GA can estimate $g_N$ by the ratio of the relative sizes of
the projected and unprojected Hilbert spaces as,
\begin{equation}
g_N\approx C \, \frac
{
(N_{\rm L}-N_{\uparrow})! \,
(N_{\rm L}-N_{\downarrow})!
}
{(N_{\rm L}-N_{\uparrow}-N_{\downarrow})!}
~,
\label{gn}
\end{equation}
where $N_{\rm L}$ is the number of lattice sites and $N_{\uparrow}$
($N_{\downarrow}$) is the number of up (down) spins.

We here discuss renormalization of spins using Eq.~(\ref{gn}).
Since $[P_{\rm G},{\bf S}_i]=0$, if a
wave function before the projection is an eigenstate of $(\sum_i{\bf
S}_i)^2$ and/or $\sum_i S^z_i $ with eigenvalues
$S(S+1),M$, respectively, then these quantities are not changed by
$P_{\rm G}$.
If it is not such an eigenstate, $P_{\rm G}$ may change the distribution of
$S,M$.  If $N$ is fixed, by changing $M$ in Eq.~(\ref{gn}), one can see
that $P_{\rm G}$ excludes more states with small $M$.  As a result, the most
``probable'' $M$ increases.
In fact, Eq.~(\ref{gn}) correctly reproduce the limit of
fully polarized states ($N_\sigma=0$),
which are obviously not affected by $P_{\rm G}$.
By rotating spin axes, and repeating this argument for the $x,y$ directions,
we conclude that $P_{\rm G}$ excludes more states with small ${\bf S}^2$.
Physically, this can be explained as follows: To make small ${\bf S}^2$,
electrons have to cancel their spin moments, and then they have high
chance to meet each other on a certain site.  When ${\bf S}^2$ is large,
the spin of an electron tend to orient the same direction as those of
the other, then electrons prefer to stay at different sites, and not
affected by the projection so much.

The Gutzwiller projection makes more singly occupied sites, and local
spin moments also tend to be magnified (this is probably related to
increase of ${\bf S}^2$).  In the canonical scheme,
magnitude of uniform (ferromagnetic) moments are restricted by the
canonical constraint, whereas non-uniform ({\it e.g.}\ antiferromagnetic,
sinusoidal) moments are free from the canonical constraint and can be
enhanced.
On the other hand, in the spin--grand-canonical scheme, the total
spin-$z$ component does not have such restriction, and ferromagnetic
moments can be also enhanced (shown in
Sec.~\ref{chargeqcano-spingcano}).

\section{Electron addition/removal matrix elements}
\label{sec:highDOS}

The general expressions of
Eqs.~(\ref{eq:cPcPhigh1}-\ref{eq:cPPchigh4}) are written as follows
using $\bar{n}_{i\sigma}\equiv 1-n_{i\sigma}$, $
\alpha_{i\sigma} \equiv { n_{i\sigma} }{ (1-n_{i\sigma})^{-1} }$ and
$\tilde{A}_{ij}\equiv n_{ji\uparrow} n_{ji\downarrow}^* + \Delta_{ji}^*
\Delta_{ij} $ $(i\neq j)$:
\begin{align}
\frac{ \langle  c_{j\uparrow}^\dagger P c_{i\uparrow} P \rangle_0 }
{ \langle  P^2 \rangle_0 }
& \approx
\sqrt{ g_{ii\uparrow}^{t0} }
\left( n_{ji\uparrow}
+
\frac{  n_{ji\downarrow} \alpha_{j\uparrow} \tilde{A}_{ij} }
{ \bar{n}_{i\downarrow}\bar{n}_{j\downarrow} }
\right)
,
\\
\frac{ \langle  c_{j\uparrow} P c_{i\uparrow}^\dagger P \rangle_0 }
{ \langle  P^2 \rangle_0 }
& \approx
\sqrt{ g_{ii\uparrow}^{t0} } 
\left( - n_{ij\uparrow}
+
\frac{ n_{ij\downarrow} \tilde{A}_{ij}^* }
{ \bar{n}_{i\downarrow}\bar{n}_{j\downarrow} }
\right)
,
\end{align}
\begin{align}
\frac{ \langle  c_{j\downarrow}^\dagger P c_{i\uparrow}^\dagger P \rangle_0 }
{ \langle  P^2 \rangle_0 }
& \approx
\sqrt{ g_{ii\uparrow}^{t0} }
\left(
- \Delta_{ij}^*
+
\frac{ \alpha_{j\downarrow} \Delta_{ji}^* \tilde{A}_{ij}^* }
{ \bar{n}_{i\downarrow}\bar{n}_{j\uparrow} }
\right)
,
\\
\frac{ \langle  c_{j\downarrow} P c_{i\uparrow} P \rangle_0 }
{ \langle  P^2 \rangle_0 }
& \approx
\sqrt{ g_{ii\uparrow}^{t0} }
\left(
 \Delta_{ij}
+
\frac{ \Delta_{ji}\tilde{A}_{ij} }
{ \bar{n}_{i\downarrow}\bar{n}_{j\uparrow} }
\right)
,
\end{align}
\begin{align}
\frac{ \langle  c_{j\uparrow}^\dagger P^2 c_{i\uparrow} \rangle_0 }
{ \langle  P^2 \rangle_0 }
& \approx
 n_{ji\uparrow}
-
\frac{  n_{ji\downarrow} \alpha_{i\uparrow}\alpha_{j\uparrow} \tilde{A}_{ij} }
{ \bar{n}_{j\downarrow}\bar{n}_{i\downarrow} }
~,
\\
\frac{ \langle  c_{j\uparrow} P^2 c_{i\uparrow}^\dagger \rangle_0 }
{ \langle  P^2 \rangle_0 }
& \approx
 - n_{ij\uparrow}
+
\frac{ n_{ij\downarrow} \tilde{A}_{ij}^* }
{ \bar{n}_{j\downarrow}\bar{n}_{i\downarrow} }
~,
\end{align}
\begin{align}
\frac{ \langle  c_{j\downarrow}^\dagger P^2 c_{i\uparrow}^\dagger \rangle_0 }
{ \langle  P^2 \rangle_0 }
& \approx
- \Delta_{ij}^*
+
\frac{ \alpha_{j\downarrow} \Delta_{ji}^* \tilde{A}_{ij}^* }
{ \bar{n}_{i\downarrow}\bar{n}_{j\uparrow} }
~,
\\
\frac{ \langle  c_{j\downarrow} P^2 c_{i\uparrow} \rangle_0 }
{ \langle  P^2 \rangle_0 }
& \approx
 \Delta_{ij}
-
\frac{ \alpha_{i\uparrow} \Delta_{ji}\tilde{A}_{ij} }
{ \bar{n}_{i\downarrow}\bar{n}_{j\uparrow} }
~.
\end{align}

\section{Higher order terms for the global constraint}
\label{sec:hiorder}

By taking up to the second order of the intersite contractions, $\langle
P^2 \rangle_0$ is represented by
\begin{equation}
\frac{\langle P^2 \rangle_0}
{\prod_{l} \Xi_l}
\approx 1 +  \sum_{l< m} \frac{\Xi^{(2)}_{lm}}{\Xi_l \Xi_m},
\label{eq:xi2}
\end{equation}
Here, $\Xi^{(2)}_{lm}$ contains all the terms of the second order of
intersite contractions in $\langle P^2 \rangle_0$.
The division by ${\prod_{l} \Xi_l}$ cancels single site
contribution and simplifies the expression.  For calculating $\langle
n_{i\uparrow} \rangle$, we need
\begin{eqnarray}
&& \frac{\langle n_{i\uparrow}P^2 \rangle_0}{\prod_l\Xi_l} =
  \frac{ \lambda (1-n_{i\downarrow}) }{\Xi_i} n_{i\uparrow}
\left (1 + \sum_{l<m,l\neq i,m\neq i}
\frac{\Xi^{(2)}_{lm}}{\Xi_l \Xi_m}
\right)
\nonumber\\&& \quad
+\sum_{l\neq i} \frac{\lambda}{\Xi_i\Xi_l}  \Big\{ 
 a_{l\downarrow}
\left[ ( 1 - n_{i\downarrow}) |n_{il\uparrow}|^2 + n_{i \uparrow} |\Delta_{li}|^2 \right]
\nonumber\\&& \qquad\qquad\quad
- a_{l \uparrow}
\left[ n_{i\uparrow} |n_{il\downarrow}|^2 + ( 1 - n_{i\downarrow} ) |\Delta_{il}|^2 \right]
\Big\}.
\label{eq:nisP}
\end{eqnarray}
By taking the ratio between Eqs.~(\ref{eq:xi2}) and (\ref{eq:nisP}), and
neglecting fourth order terms, contribution from disconnected clusters
disappears.  Then, we obtain renormalization of particle densities as
Eq.~(\ref{eq:globalnrenorm}).

To determine $\lambda^{(2)}$, we use the equation for
$1-\langle \hat{n}_{i\uparrow}+\hat{n}_{i\downarrow} \rangle $, namely,
\begin{eqnarray}
&& N_{\rm L} - N^{\rm after} =
\sum_i  \frac{ (1-n_{i\uparrow})(1-n_{i\downarrow}) }{\Xi_i}
 \nonumber\\ && \qquad
+ \sum_i \lambda \frac{ n_{i\uparrow}(1-n_{i\downarrow})
               +n_{i\downarrow}(1-n_{i\uparrow})}{\Xi_i}
 \sum_{l\neq i}
\frac{\Xi^{(2)}_{il}}{\Xi_{i} \Xi_{l}}
 \nonumber\\ && \quad
-\sum_i\sum_{l\neq i} \frac{\lambda}{\Xi_i\Xi_l}
\times
 \nonumber\\ && \quad
\Big\{ 
 a_{l\downarrow}
\left[
( 1 - 2n_{i\downarrow}) |n_{il\uparrow}|^2  -
(1-2n_{i \uparrow}) |\Delta_{li}|^2
\right]
\nonumber\\&& \quad
+ a_{l \uparrow}
\left[
( 1 - 2n_{i\uparrow}) |n_{il\downarrow}|^2 -
( 1 - 2n_{i\downarrow} ) |\Delta_{il}|^2
\right]
\Big\}.
\end{eqnarray}
By replacing $\lambda$ by $\lambda^{(0)}+\lambda^{(2)}$, zeroth order
term cancels between the l.h.s.\ and the first term of the r.h.s.  The
later also contains $\lambda^{(2)}$; in fact, $\lambda^{(2)}$ in the other
terms can be negligible because they are multiplied to other intersite
contractions.
Regarding $\lambda^{(2)}$ as the same order as $n_{ij\sigma}^2$ and
$\Delta_{ij}^2$, and neglecting high order terms, Eq.~(\ref{eq:lambda2})
is obtained.

\end{document}